\documentclass[iop]{emulateapj}
\usepackage{epsfig}
\usepackage{csquotes}
\usepackage{graphicx}
\usepackage{ulem,color}
\usepackage[dvipsnames]{xcolor/xcolor}
\usepackage{dcolumn}
\usepackage{bm}
\usepackage{longtable}
\usepackage{natbib}
\usepackage{hyperref}
\usepackage{xspace}
\usepackage{amsmath}
\hypersetup{linkcolor=red,citecolor=orange,filecolor=cyan,urlcolor=magenta}%

\def\lesssim{\mathrel{\hbox{\rlap{\hbox{\lower4pt\hbox{$\sim$}}}\hbox{$<$}}}}
\def\gtrsim{\mathrel{\hbox{\rlap{\hbox{\lower4pt\hbox{$\sim$}}}\hbox{$>$}}}}
\newcommand{\bea}{\begin{eqnarray}}
\newcommand{\eea}{\end{eqnarray}}

\newcommand{\bP}{{\bf P}}
\newcommand{\bF}{{\bf F}}
\newcommand{\bU}{{\bf U}}

\newcommand{\prim}{{{\mathbf{P}}}}

\newcommand{\harm}{{\sc Harm3d}\xspace}

\newcommand{\ssim}{\texttt{S06}}
\newcommand{\nossim}{\texttt{S0}}

\def\lambdabar{%
\relax
\bgroup
\def\@tempa{\hbox{\raise.73\ht0
\hbox to0pt{\kern.25\wd0\vrule width.5\wd0
height.1pt depth.1pt\hss}\box0}}%
\mathchoice{\setbox0\hbox{$\displaystyle\lambda$}\@tempa}%
{\setbox0\hbox{$\textstyle\lambda$}\@tempa}%
{\setbox0\hbox{$\scriptstyle\lambda$}\@tempa}%
{\setbox0\hbox{$\scriptscriptstyle\lambda$}\@tempa}%
\egroup
}

\usepackage{soul}

\providecommand{\figfolder}{./}
\DeclareRobustCommand{\firstrev}[1]{ {\begingroup\rm{#1}\endgroup} }

\DeclareRobustCommand{\secondrev}[1]{ {\begingroup\rm{#1}\endgroup} }

\slugcomment{Accepted for publication in ApJ}

\begin{document}

\title{Mini-disk accretion onto spinning black hole binaries:\\ quasi-periodicities and outflows}

\author{
  Luciano Combi    $^{1,2}$,
  Federico G. L\'opez Armengol $^2$,
  Manuela Campanelli $^2$,
  Scott C. Noble $^{3}$,
  Mark Avara $^{4}$,
  Julian H. Krolik $^5$
  Dennis Bowen $^{6,7}$
} 

\affil{$^1$ Instituto Argentino de Radioastronom\'ia (IAR, CCT La Plata, CONICET/CIC), C.C.5, (1984)
  Villa Elisa, Buenos Aires, Argentina\\
  $^2$ Center for Computational Relativity and Gravitation,
  Rochester Institute of Technology, Rochester, NY 14623\\
  $^3$ Gravitational Astrophysics Laboratory, NASA Goddard Space Flight Center, Greenbelt, MD 20771.\\
  $^4$ Institute of Astronomy, University of Cambridge Madingley Road, CB3 0HA Cambridge, UK\\
  $^5$ Department of Physics and Astronomy, Johns Hopkins
  University, Baltimore, MD 21218\\
  $^6$ Center for Theoretical Astrophysics, Los Alamos National Laboratory,
  P.O. Box 1663, Los Alamos, NM 87545.\\
  $^7$ X Computational Physics, Los Alamos National Laboratory,
  P.O. Box 1663, Los Alamos, NM 87545.
}

\email{lcombi@iar.unlp.edu.ar}

\begin{abstract}
We perform a full 3D general relativistic magnetohydrodynamical (GRMHD) simulation of an equal-mass, spinning, binary black hole approaching merger, surrounded by a circumbinary disk and with mini-disks around each black hole.  For this purpose, we evolve the ideal GRMHD equations on top of an approximated spacetime for the binary that is valid in every position of space, including the black hole horizons, during the inspiral regime. We use relaxed initial data for the circumbinary disk from a previous long-term simulation, where the accretion is dominated by an $m=1$ overdensity called the lump. We compare our new spinning simulation with a previous non-spinning run, studying how spin influences the mini-disk properties. We analyze the accretion from the inner edge of the lump to the black hole, focusing on the angular momentum budget of the fluid around the mini-disks. We find that mini-disks in the spinning case have more mass over a cycle than the non-spinning case. However, in both cases we find most of the mass received by the black holes is delivered by direct plunging of material from the lump. We also analyze the morphology and variability of the electromagnetic fluxes and we find they share the same periodicities of the accretion rate. In the spinning case, we find that the outflows are stronger than the non-spinning case. Our results will be useful to understand and produce realistic synthetic light curves and spectra, which can be used in future observations.
\end{abstract}


\section{Introduction}
\label{sec:introduction}

When two galaxies merge, a supermassive binary black hole (SMBBH) is expected to form \citep{Merritt2005}. The potential interaction of the new system with the surrounding gas and the dynamical friction of stars might shrink the binary separation to sub-parsec scales \citep{Begelman1980,
Mayer2007, Escala2004, Escala2005, Merrit2004, Merritt2006, Dotti2007,
Dotti2009b, Shi2012,SesanaKhan2015,Mirza2017,Khan2019, Tiede+2020, dittmann2022}. At those separations, energy and angular momentum is extracted from the system by gravitational radiation until the BHs merge \citep{PhysRevLett.95.121101,PhysRevLett.96.111101, PhysRevLett.96.111102}. In the near future, gravitational waves from SMBBH mergers might be observable in the mHZ frequency band by the \textit{Laser Interferometer Space Antenna} \citep[LISA,][]{LISA2017} and by \textit{pulsar timing} techniques in
the nHz range \citep{Nanograv2020, EPTA2016, Reardon2015}. 

Since the environment of these systems is most likely gas-rich, a SMBBH could also emit electromagnetic radiation through accretion \citep{Barnes1992, Barnes1996, Mihos1996,Mayer2007, Dotti2012, Mayer2013, Derdzinski2019}. At subparsec scales, SMBBHs cannot be spatially resolved and they might be hard to distinguish from ordinary active galactic nuclei (AGN). There are many proposed signatures to identify the presence of a SMBBH using electromagnetic waves, for instance: Doppler variations due to the orbital motion \citep{dorazio2015nature}, binary periodicities in the light curves \citep{Valtonen2006, Graham2015a,Graham2015b, Liu2019,Saade2020}, interruption of jet emission \citep{Shoenmakers2000, Liu2003} and “spin-flips” of the BH after merger \citep{Merritt2002}, dual-radio cores \citep{Rodriguez2006}, profile shifts of broad emission lines \citep{Dotti2009, Bogdanovic2009}, a ``notch” in the optical/IR spectrum \citep{Roedig2014, Sesana2012}, periodicities in the thermal spectrum due to a short residence time for gas in the disk \citep{Bowen2019} and X-ray periodicities \citep{Sesana2012,Roedig2014}. The feasibility of detecting some of these signatures depends strongly on binary properties such as mass-ratio and orbital separation (see also \cite{Krolik2019} for likely source counts). 

Because the interstellar gas of the merged galaxy would have a considerable amount of angular momentum, a circumbinary disk should form around the binary \citep{Springel2005, Chapon2013}. For mass-ratios close to one, the system would present a gap of radius $\sim 2a$ between the binary's semimajor axis $a$ and the edge of the circumbinary disk. Accretion then occurs through two streams from the circumbinary to each black hole. Depending on its angular momentum, the material from the streams can eventually start orbiting the BHs, forming ``mini-disks''. On the other hand, the time-dependent quadrupole potential of the binary system can induce a density concentration in the edge of the circumbinary disk on a small azimuthal range, breaking the axisymmetric accretion \citep{MM08}. This feature is usually referred as the \textit{lump} \citep{Shi12, Noble12, noble2021}. When the lump is formed, it behaves as a coherent $m=1$ density mode orbiting the system at a frequency $\sim 0.25 \: \Omega_{\rm bin}$, where $\Omega_{\rm bin}$ is the binary frequency \citep{noble2021, armengol2021}. The time-dependence of the gas available for accretion onto the black holes is then dominated by the lump. If the residence time of the gas in the mini-disk is shorter (or comparable to) the modulation period of the lump, then the accretion rate and luminosity of the mini-disks are modulated by the lump  \citep{Bowen2018, dAscoli2018, Bowen2019}. Whether this modulation happens depends mostly on the orbital separation (see Section \ref{sec-results})

Numerical simulations are key tools to make accurate models from these highly non-linear systems. Circumbinary accretion has been largely investigated using viscous hydrodynamical models in 2D with Newtonian gravity \citep{MM08,DOrazio13,Farris2014a,Farris2014b,DOrazio2016,Munoz2016,Miranda2017,
Derdzinski2019,Munoz2019,Moody2019,Mosta2019,Duffell2020,Zrake2020,Munoz2020a,Munoz2020b,
Tiede2020,Derdzinski2021}. In these simulations, the system can be evolved for $\mathcal{O}(100-1000)$ binary orbits, when a relaxed stage is reached, and the effects of the initial conditions are suppressed. However, these simulations rely on artificial sink conditions for the black hole and their stress model is not self-consistent, i.e. they adopt an ad-hoc description for the viscosity.

More realistic simulations using MHD have been performed in Newtonian gravity \citep{Shi12,Shi2015}, in approximated GR \citep{Noble12, Zilhao2015, armengol2021, noble2021, Bowen2018, Bowen2019},  and in full numerical relativity \citep{Gold14, Gold2014b, Farris11, paschalidis2021minidisk,cattorini2021,giacomazzo2012}. MHD simulations are more computationally expensive than two-dimensional $\alpha-$viscous simulations because they demand three-dimensional domains and fine resolutions. Moreover, GRMHD simulations including both the mini-disks and the circumbinary disk need to resolve different dynamical timescales, evolve the spacetime metric, and thus are usually too expensive to evolve for many orbits.

Since the properties of the mini-disks are necessarily tied to the circumbinary accretion, it is important to perform simulations linking these regimes in order to obtain a proper global description of the system. A viable method to accomplish this was presented in \cite{Bowen2018}, where a snapshot of an evolved circumbinary disk simulation \citep{Noble12} was used as initial data for studying mini-disk accretion onto non-spinning black holes. In this way, \cite{Bowen2019} showed that mini-disks at close binary separations exhibit a quasi-periodic filling and depletion cycle determined by the lump and the short inflow time of the mini-disks.

An important property of supermassive binary black holes that many of these studies miss is the spin \citep{Lin1979}. When binary BHs are approaching merger, spins can have an important impact on the spacetime evolution: they can alter the orbital motion of the system \citep{Campanelli:2006uy,  hemberger2013final,healy2018hangup}, induce precession and nutation \citep{Campanelli:2006fy}, repeatedly flip their sign \citep{Lousto:2014ida, Lousto:2015uwa, lousto2016spin, Lousto:2018dgd} and even tilt the orbital orientation \citep{kesden2014gravitational}. Spins also have a key role in the accretion of matter into BHs \citep{KHH05} and binary BHs. For instance, accretion rate per unit mass near the circumbinary disk's inner edge depends on the spin, altering the mass profile in the inner part of the disk \citep{armengol2021}. Moreover, the character of the flow within the mini-disks depends on the ratio between the radius at which they are fed and the radius of the innermost stable circular orbit (ISCO) \citep{Bowen2018, Gold14, paschalidis2021minidisk}, which is strongly dependent on the spin. 

On the other hand, spinning black holes are expected to launch electromagnetic outflows \citep{dVH03, HKdVH04, mckinney2004measurement, dVHKH05, KHH05, TchekhovskoyChap}. The generation of a net Poynting flux in a binary BH system approaching merger has been modeled within General Relativistic Force-Free Electrodynamics (GRFFE) \citep{palenzuela2010magnetospheres, palenzuela2010dual, neilsen2011, moesta2012detectability} and ideal GRMHD \citep{Giacomazzo12,Kelly2017, Kelly2021,Farris12,Gold:2013APS,Gold14,paschalidis2021minidisk}. In the force-free regime, given a homogeneous plasma threaded by a constant magnetic field, the Blandford-Znajek (BZ) mechanism \citep{blanford1977} proved to operate efficiently around each BH, leading to a pair of collimated jets in a double-helical structure that coalesces after merger \citep{palenzuela2010magnetospheres}. The dynamics of binary BHs in a homogeneous medium have also been investigated in the ideal GRMHD regime, where the inertia of the plasma is taken into account; in this case, accretion of the plasma onto the BHs leads to a steeper growth of the initial magnetic field during the inspiral, via compression and magnetic winding, resulting in higher luminosities \citep{Giacomazzo:2012ApJ}. 
At the same time, the inertia of the (homogenous) plasma falling onto the BHs interferes with the propagation of the electromagnetic flux, causing less collimated magnetic structures compared with GRFFE  \citep[see also][]{Kelly2017, Kelly2021}.  Further, \cite{Farris12,Gold:2013APS,Gold14,paschalidis2021minidisk} performed GRMHD simulations that include the circumbinary disk in the domain and found the development of net Poynting fluxes emitted from the polar regions of each BH that coalesce at larger distances.

There are only a few GRMHD simulations of circumbinary accretion into spinning SMBBHs. Very recently, \cite{armengol2021} presented the first long-term circumbinary GRMHD simulation of spinning black holes (with the inner cavity excised), and \cite{paschalidis2021minidisk} presented the first numerical relativity simulation of mini-disks around spinning BHs, confirming previous expectations that the ISCO plays a key role in the mini-disk mass \citep{Bowen2018,Bowen2019,Gold2014b}. A natural next step is to consider more realistic simulations where both mini-disks and a properly relaxed circumbinary disk around spinning BBHs are taken into account. This is important to make accurate predictions of the light curves and spectra of these systems.

In this work, we present the results of a GRMHD simulation of mini-disk accretion around spinning black holes of spins $a = 0.6~M$ aligned with the orbital angular momentum. We evolve the ideal GRMHD equations on top of a binary BH spacetime that is moving on a quasi-circular orbit starting at $20 M$ separation. We use an approximate BBH spacetime that uses Post-Newtonian (PN) trajectories for the BHs in the inspiral regime, but is valid at every space point, including the BH horizons. As initial data for the plasma, we use a steady-state snapshot of a circumbinary disk simulation performed in \cite{Noble12}. We compare our new simulation with a previous non-spinning simulation \citep{Bowen2018,Bowen2019} that uses the same initial data. 

The paper is structured as follows: In Section \ref{sec-simusetup}, we present the simulations set-up, the numerical methods, the spacetime approximation that we use, and the initial data. In Section \ref{sec-results}, we present our results. First, we give an overview of the main features of the system and the relation with previous simulations. In Section \ref{sec-mdot} we investigate the accretion rate, inflow time, and mass evolution of the mini-disks, and in Section \ref{sec-struct} we analyze the mini-disks structure, the specific angular momentum distribution, and the azimuthal density modes. Then, in Section \ref{sec-emfluxes} we analyze the outflows, the magnetized structure of the system, and the variability of the Poynting flux, comparing spinning and non-spinning results. Finally, in Section \ref{sec-discussion} we discuss some of the implications of our results and in Section \ref{sec-conclusions} we summarize the main points and results of the paper.

\textbf{Notation and conventions.} We use the signature $(-,+,+,+)$ and we follow the Misner-Thorne-Wheeler convention for tensor signs. We use geometrized units, $G=c=1$. We use Latin letters $a,b,c,...=0,1,2,3$ for four dimensional components of tensors, and $i,j,k,...=1,2,3$ for space components.

\section{Simulation set-up}
\label{sec-simusetup}

We evolve the equations of ideal GRMHD in the spacetime of a binary black hole system using the finite-volume code \harm. Our goal is to analyze the effects of the black hole spin in the minidisks. For that purpose, we compare two simulations of an equal-mass binary black hole, with and without spins. The non-spinning simulation, denoted as \nossim{}, was performed in \cite{Bowen2019} using an analytical metric built by matching different spacetimes \citep{Mundim2014}. We perform a new simulation, denoted \ssim{}, with BHs having spins of $\chi=0.6$ aligned with the orbital angular momentum of the binary, using an approximate analytical metric described in \cite{combi2021} (a concise summary is given in Sec.~\ref{sec:metric}). For both simulations, we use the same grid and initial data in order to have a faithful comparison (see Section~\ref{sec:metric} for details). 

\subsection{GRMHD equations of motion}
\label{sec:EOM}

Assuming that the plasma does not influence the spacetime, we evolve the ideal GRMHD equations in the dynamical binary BH metric described in Section \ref{sec:metric}. The equations of motion are given by the conservation of the baryon number, the conservation of the energy-momentum tensor, and Maxwell's equations with the ideal MHD condition:
\begin{equation*}
\nabla_a (\rho u^a) = 0, \quad \nabla_{a} {T^{a}}_{b} = \mathcal{F}_{b}, 
\end{equation*}
\begin{equation}
\nabla_{a} \:^{*} F^{ab} = 0,  \quad u_{a} F^{ab} =0,
\end{equation}
where $\rho$ is the rest-mass density, $F^{ab}/\sqrt{4 \pi}$ is the Faraday tensor\footnote{Following \cite{Noble09}, we absorb the factor $1/\sqrt{4 \pi}$ in the definition of the tensor $F^{ab}$.}, $u^{a}$ is the four-velocity of the fluid, $\mathcal{F}_b$ is the radiated energy-momentum per 4-volume unit, and the MHD energy-momentum tensor is
\begin{equation}
T^{ab} := (\rho h + b^2) u^a u^b + (P + \frac{1}{2} b^2) g^{ab} - b^a b^b,
\end{equation}
where $h:= (1+ \epsilon +P/\rho)$ is the specfic enthalpy, $\epsilon$ is the specific internal energy, $P$ is the pressure, $b^a:= \:^{*}F^{ab}u_b:$ is the four-vector magnetic field, and $b^2:=b^ab_a$ is proportional to the magnetic pressure, $p_{\rm m} := b^2/2$. We follow \cite{Noble12,Noble09} and write these coupled equations of motion in manifest conservative form as
\begin{equation}
\partial_t \bU\left(\prim\right) =                                                                       
-\partial_i \bF^i\left(\prim\right) + \mathbf{S}\left(\prim\right),
\label{eq:cons-form-mhd}
\end{equation}
where $\bP$ are the \textit{primitive} variables, $\bU$ the \textit{conserved} variables, $\bF^i$ the \textit{fluxes}, and $\mathbf{S}$ the \textit{source} terms.  These are given explicitly as:
\begin{eqnarray}
\mathbf{P}: & = & [\rho, u, \tilde{u}^j, B^{j}] \ ,
\label{primitive-mhd} \\
\bU\left(\prim\right) & = & \sqrt{-g} \left[ \rho u^t ,\, {T^t}_t +
  \rho u^t ,\, {T^t}_j ,\, B^j\right] \ , \label{cons-U-mhd} \\
\bF^i\left(\prim\right) & = & \sqrt{-g} \left[ \rho u^i ,\, {T^i}_t +
  \rho u^i ,\, {T^i}_j ,\, \left(b^i u^j - b^j u^i \right)\right], \label{cons-flux-mhd} \\
\mathbf{S}\left(\prim\right) & = & \sqrt{-g} \left[ 0 ,\,
  {T^\kappa}_\lambda {\Gamma^\lambda}_{t \kappa} - \mathcal{F}_t ,\,
  {T^\kappa}_\lambda {\Gamma^\lambda}_{j \kappa} - \mathcal{F}_j ,\, 0
  \right], \label{cons-source-mhd}
\end{eqnarray}    
where $g$ is the determinant of the metric, ${\Gamma^\lambda}_{\alpha \beta}$ 
are the Christoffel symbols, $u:=\rho \epsilon$ is the internal energy, $\tilde{u}^j:= u^j -g^{tj}/g^{tt}$ is the velocity relative to the normal spacelike hypersurface, and $B^{j}:= \: ^{*} F^{it}$ is the magnetic field, which is both a conserved and a primitive variable \footnote{We denote the magnetic field in the frame of normal observers (proportional to the constrained-transported field) as $B^i$, while we denote the magnetic field in the frame of the fluid as $b^{a} = \left(1/u^t\right)\left({\delta^a}_\nu + u^a u_b \right)B^b$.}. We close the system with a $\Gamma$-law equation of state, $P=(\Gamma-1) \rho \epsilon$, where we set $\Gamma= 5/3$.

The source term in the energy-momentum conservation ensures that part of the dissipated energy caused by MHD turbulence is converted to radiation that escapes from system. We assume radiation is removed from each cell independently of all the others, isotropically in the fluid frame. In this way, we set $\mathcal{F}_{a} = {\cal L}_{c}u_\beta$, where $\mathcal{L}_c$ is the cooling function. We use the prescription used in \cite{Noble12} for the rest-frame cooling rate per unit volume:
\begin{equation}
  \mathcal{L}_{c} = \frac{\rho \epsilon}{t_{\mathrm{cool}}} \left( \frac{\Delta S}{S_0} + \left| \frac{\Delta S}{S_0}\right| \right)^{1/2} \ , 
\end{equation}
where $t_{\mathrm{cool}}$ is the cooling timescale, where the disk radiates away any local increase in entropy, $\Delta S := S - S_0$, where $S := p / \rho^{\Gamma}$ is the local entropy. Our target entropy, $S_0 = 0.01$, is the initial entropy of each accretion disk in the simulation. The timescale $t_{\mathrm{cool}}$ is determined by the local fluid orbital period, following the prescriptions in \cite{Bowen17,dAscoli2018}.

\begin{center}
\begin{deluxetable}{c c c}[ht!]
\tablewidth{\columnwidth}
\tablecolumns{3}
\tablecaption{\label{tab:grid}}
\tablehead
{
  \colhead{Simulation}  & \colhead{\texttt{S06}}     & \colhead{\texttt{S0}}
}
\startdata
Spin parameter [$\chi$]   	 		     &  0.6    &  0.0       \\
BH$_{1,2}$ mass [$M_{\rm BH}$] 		         &    0.5    & 			\\
Mass-ratio [$q$] 		         &    1    & 			\\
Final time [$t_f$]               & $8000 \: M$ & $6000 \: M$ \\
Final separation [$r_{12}(t_f)$] & $16.6 \: M$ & $17.8 \: M$ \\
\# Orbits 						 & $15$	   & $12.5$    \\
Init. separation [$r_{12}(0)$]   &  $20 \: M$      &    \\
Init. total mini-disk mass [$M_0$] & $20$  &   \\
Average orbital period [$T_{\mathcal{B}}$] &  & $530\: M$   \\
Lump orbital frequency [$\Omega_{\rm lump}$] & $0.28 \:  \Omega_{\mathcal{B}}$ &    \\
ISCO radius [$r_{\rm ISCO}$] & $2.82 \:  M_{\rm BH}$ & $5.0 \:  M_{\rm BH}$   \\
Truncation radius [$r_{\rm trunc}$] & $0.4 \: r_{12}$ &    \\

\hline \\
Grid [$(x^1 \times x^2 \times x^3)$] &      & $(600,160,640)$ \\
Physical Size [($r_{\rm min},r_{\rm max}$)]   &  & $(2M,260M)$ \\ 
\enddata
\tablecomments{Physical and grid parameters of both non-spinning and spinning simulations.} 
\end{deluxetable}
\end{center}

\subsection{Code details, grid, and boundary conditions}

We solve equations \eqref{eq:cons-form-mhd} using the high-resolution, shock-capturing methods implemented in \harm. Following \cite{Noble12}, we use a piecewise parabolic reconstruction of the primitive variables for the local Lax-Friedrichs flux at each cell interface and the Flux CT method to maintain the solenoidal constraint \citep{toth2000}. Once the numerical fluxes are found, the equations are evolved in time using the method of lines with a second-order Runge-Kutta method. The primitive variables are recovered from the evolved conserved variables using the 2D method developed in \cite{Noble06}.

\begin{figure}[htb!]
  \centering
  \includegraphics[width=\columnwidth]{\figfolder/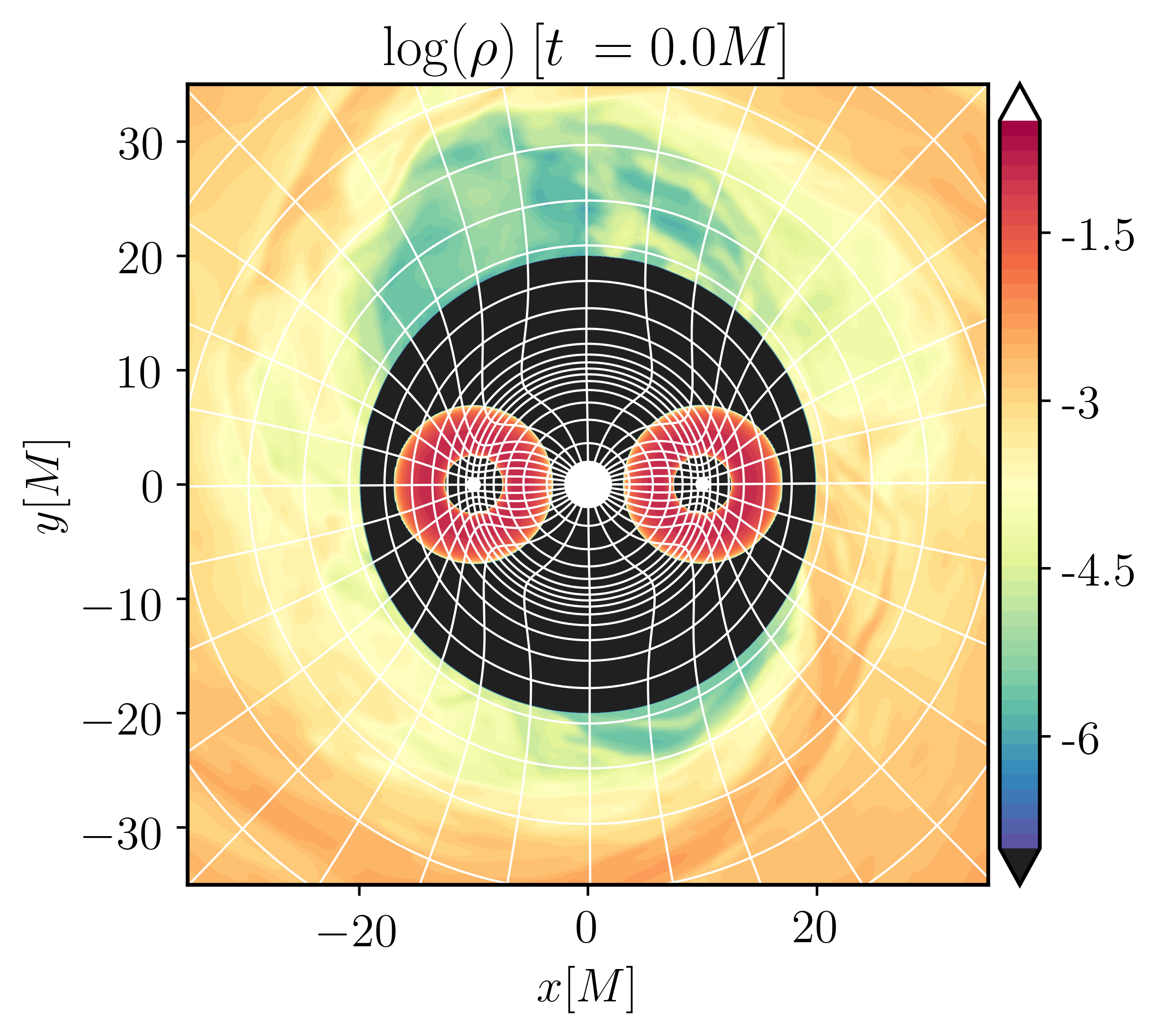}
  \caption{Initial data used in the simulation with quasi-equilibrated mini-toris around the black holes. In white thin lines we plot the warped spherical grid every each 50 cells.}
  \label{fig-initdata}
\end{figure}

\begin{figure*}[htb!]
  \centering
  \includegraphics[width=1.0\columnwidth]{\figfolder/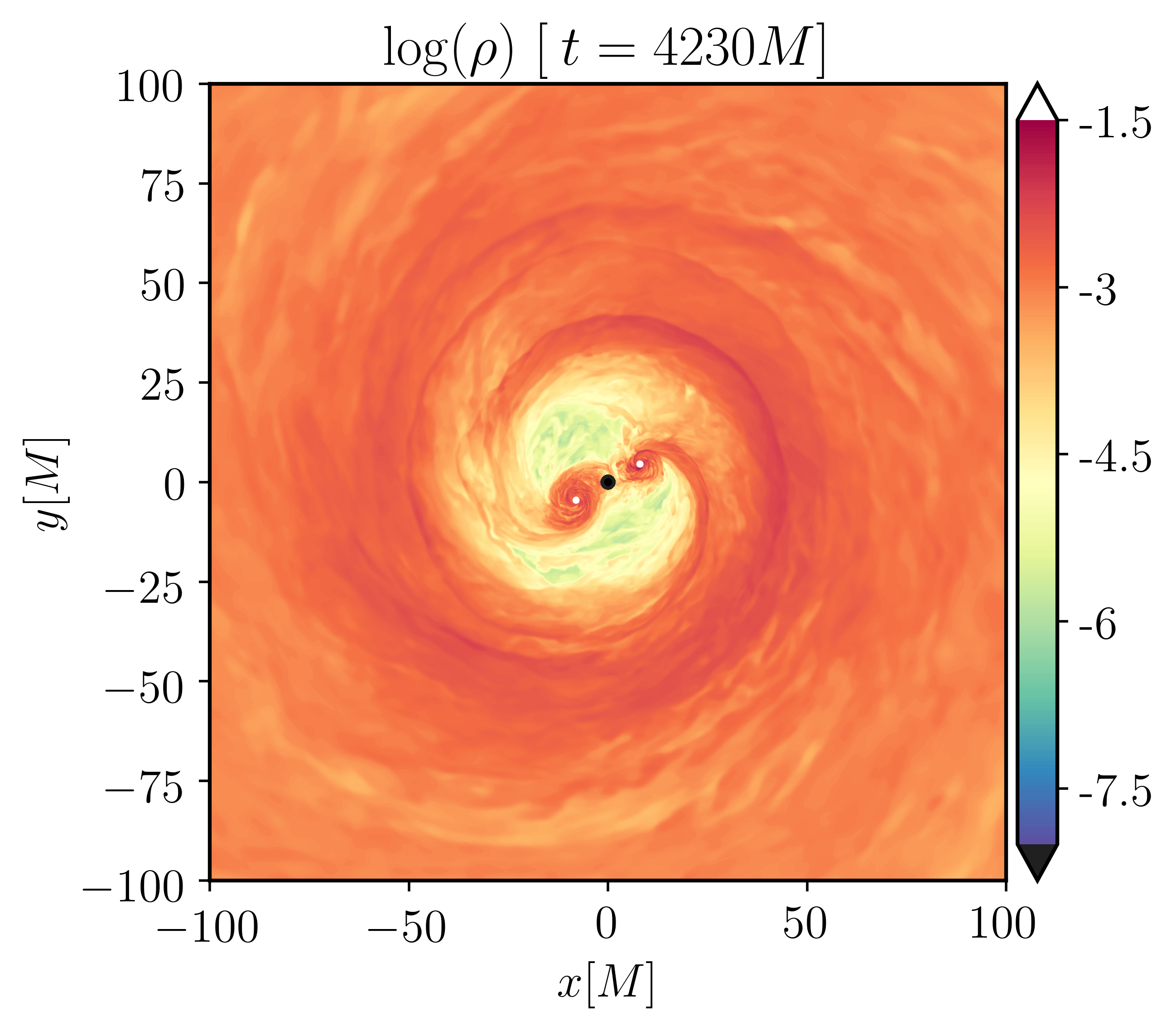}
  \includegraphics[width=\columnwidth]{\figfolder/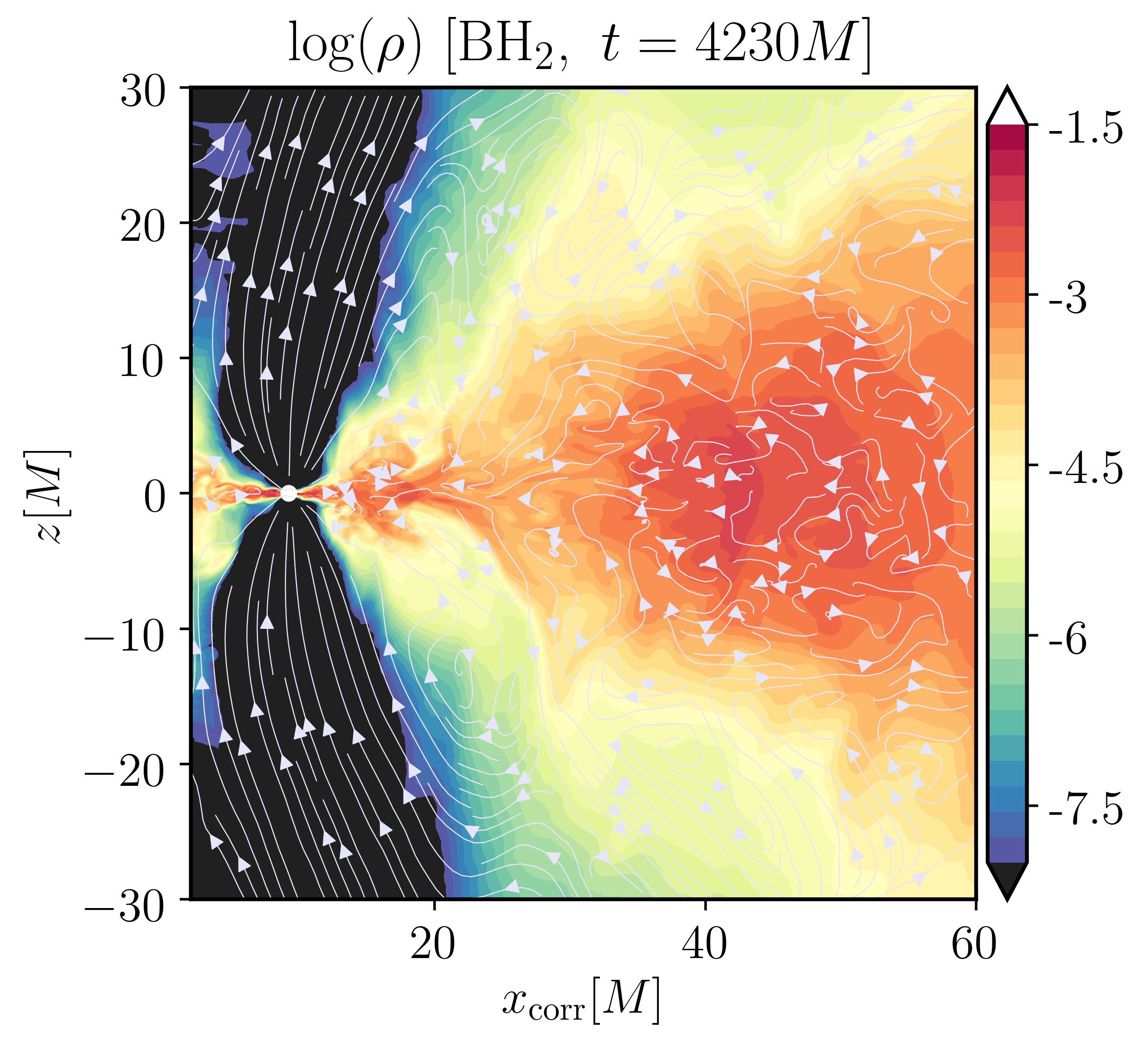}
  \caption{ Left Panel: Rest-mass density snapshot of the fluid in the equatorial plane for \ssim{}. Right panel: Rest-mass density snapshot of the fluid in the meridional plane corrotating with the (second) black hole. White stream lines represent the comoving magnetic field projected on the meridional plane.}
  \label{fig-rhp}
\end{figure*}

The grid and boundary conditions used in this simulation are the same as in \cite{Bowen2018} and \cite{Bowen2019}. We use a time-dependent, double fish-eye, warped spherical grid, centered in the center of mass of the binary system, developed in \cite{WARPED} (see full details of the grid used in \cite{Bowen2019}). The maximum physical size of the grid is set to $r_{\rm max}=13 \: r_{12}(0)$, containing the circumbinary disk of \texttt{RunSE} in \cite{Noble12}, that we use as initial data. We use outflow boundary conditions on the radial ($x^1$) boundaries, demanding the physical radial velocity $u^r$ to be oriented out of the domain; if not, we reset the radial velocity to zero and solve for the remaining velocity components. Poloidal coordinates ($x^2$) have reflective, axisymmetric boundary conditions at the polar axis cutout and the azimuthal coordinates ($x^3$) have periodic boundary conditions.

The resolution is given by $N_{r} \times N_{\theta} \times N_{\phi} = 600{\times}160{\times}640$ cells. The shape of the grid in the circumbinary region matches the grid used in \cite{Noble12} and is sufficient to resolve the magnetorotational instability (MRI) in the circumbinary disk. Because of our polar grid resolution and off-grid-center location of the BHs in the spherical grid, our configuration does not include a full 32 cells per scale height in the mini-disks on the side farthest from the center-of-mass. 

\subsection{Binary BH spacetime and initial data}
\label{sec:metric}

The spacetime of the binary black hole is approximated by superposing two Kerr spacetimes on a Minkowski background.  We describe it in terms of harmonic coordinates.  The metric can be written schematically as
\begin{equation}
g_{ab} = \eta_{ab} + M_1 \mathcal{H}^1_{ab} (\vec{x}_1,\vec{v}_1)+M_2 \mathcal{H}^2_{ab} (\vec{x}_2,\vec{v}_2),
\label{eq-supmetric}
\end{equation}
where $\eta_{ab}$ is the Cartesian Minkowski metric, $\mathcal{H}^A_{ab}$ is the boosted black hole term for $A=1,2$, $M_A$ is the mass, and $\lbrace \vec{x}_A(t), \vec{v}_A(t) \rbrace$ are the position and velocity of the black hole. The trajectories are obtained by solving the Post-Newtonian equations of motion for a spinning binary BH in quasi-circular motion at 3.5 PN order. In \cite{combi2021}, we showed that this analytical metric constitutes a good approximation to a vacuum solution of Einstein's field equations for a binary approaching merger\secondrev{, see also \cite{east2012}, \cite{varma2018} and \cite{ma2021extending} for similar approaches in the context of numerical relavity}. The metric \eqref{eq-supmetric} is computationally efficient, compared with previous approaches, and easy to handle for different parameters. In the non-spinning simulation, the spacetime was represented by an semi-analytical metric built by stitching different approximate solutions of Einstein's equation \citep{Mundim2014}. \firstrev{In \cite{combi2021}, we compared the matching and superposed metrics evolving two GRMHD simulations for the non-spinning case and we found that they are completely equivalent in this context. Moreover, we analyze the spacetime scalars and integrated Hamiltonian constraints for each one and we found that (a) they remain small and well-behaved up to a separation of $10 ~M$, and (b) the constraints remain invariant when we change the BH spin, and thus no pathologies are introduced by the spin}.

\secondrev{
The level at which these constraints are violated is comparable to the very low level achieved in the numerical simulations performed in \cite{zlochower2016inspiraling}, who used the matching metric as initial data for evolving Einstein's equations. Constraint violations in this
evolution, which can be damped using the CCZ4 scheme, mainly introduce deviations to the trajectories (eccentricity) and errors in the masses and spins of the BHs. In our analytical metric, however, there are no such dynamical effects since we solve the trajectories using the PN approximation and the BH masses/spins are fixed. The small constraint violations, relative to the mass of the BHs, might produce small errors in the gas dynamics, but these are washed out by MHD turbulence in our simulation \citep{Zilhao2015}.}

Although the metric uses PN trajectories and thus is only valid in the inspiral regime, it is mathematically well-defined at every point in space, including the horizons of the black holes; hence, no artificial sink terms or large excisions are needed in the evolution. We apply, however, a mask inside the horizon to avoid the singularity of each black hole. In particular, we do not evolve the hydro fluxes inside the masked region and we set to zero the magnetic fluxes. This allows us to evolve the induction equations in the whole domain and preserve the solenoidal constraints.

\begin{figure}
  \includegraphics[width=\columnwidth]{\figfolder/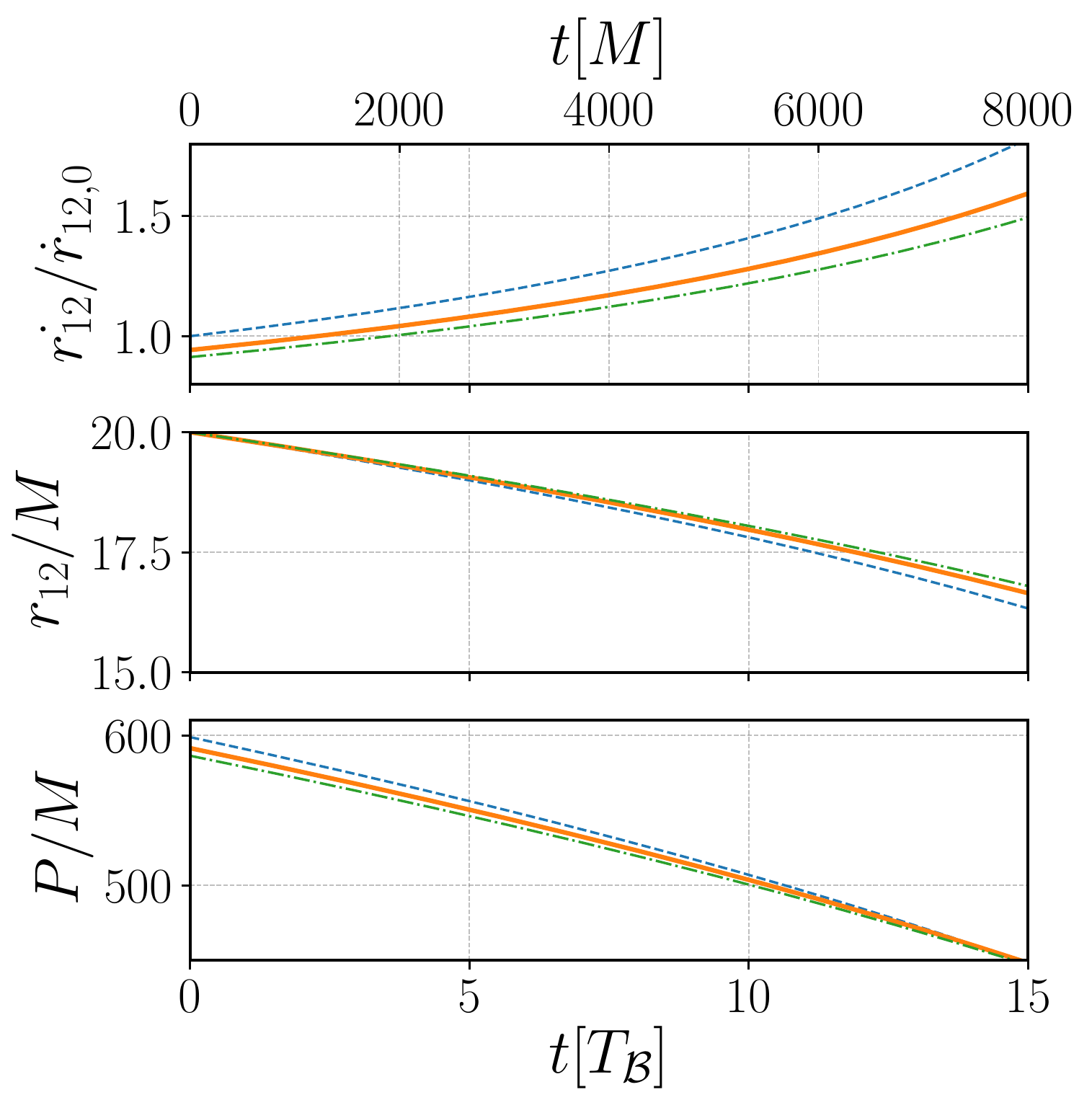}
  \caption{Properties a binary black hole system with aligned spins of $\chi=0.0$ (blue dashed line), $\chi=0.6$ (solid orange line), and $\chi=0.9$ (green dashed line). In the top panel, we plot the radial velocity, $\dot{r}_{12}$,  normalized with its initial value, in the middle panel the orbital separation $r_{12}$, and in the bottom panel the period of the orbit, $P$.}
  \label{fig-traj}
\end{figure}

For this simulation, we use an equal-mass black hole binary, $M_1=M_2=M/2$, with an initial separation of $r_{12}(0)=20\:M$, where relativistic effects are important \citep{Bowen17} and the orbit is shrinking due to gravitational radiation. In \ssim{} we set the spins of the black holes perpendicular to the orbital plane (i.e. no precession) with a moderate value of $\chi_i \equiv a_i/M = 0.6$. Since spin couples with the orbital motion, the trajectories of the holes change with respect to a non-spinning system. In particular, the inspiral is delayed because of the \textit{hang-up} effect \citep{Campanelli:2006fg}, and the orbital frequency increases (see Figure \ref{fig-traj}). 

\firstrev{In order to compare with the previous simulation, we take the same initial data for the matter fields as in \cite{Bowen2018}}. We start with a snapshot of the circumbinary disk from \cite{Noble12}, previously evolved for $5 \times 10^4 M$ ($\sim 80$ orbits). At this time, the disk is in a turbulent state and accretion into the cavity is dominated by a $m=1$ density mode, the so-called lump, that is orbiting at the inner edge of the circumbinary disk. In \cite{Noble12}, a zero spin PN metric in harmonic coordinates was used to evolve the system. As shown in \cite{armengol2021}, the bulk properties of the circumbinary disk are not sensitive to the spin, even for high values, so it is a good initial state for our spinning simulation. We interpolate this data onto our grid and we initialize two mini-disks inside the cavity, see Figure \ref{fig-initdata}. We then clean magnetic divergences introduced by the interpolation to the new grid using a projection method as explained in \cite{Bowen2018}.

\subsection{Diagnostics}

Properties of the circumbinary disk and other global properties of the system are better analyzed in the center of mass coordinates. On the other hand, to analyze mini-disk properties such as fluxes, we compute quantities on the (moving) BH frame. We define the BH frame at a given time slice with a boosted coordinate system centered at each BH (see \cite{combi2021}); we denote these BH coordinates with a bar, $\lbrace \bar{t}, \bar{r}, \bar{\theta}, \bar{\phi} \rbrace$. \firstrev{We notices that all our diagnostic are written in the harmonic coordinate gauge}. Fluxes and other local properties in this frame are computed in post-process, interpolating the global grid into a spherical grid centered in the BH with the Python package \texttt{naturalneighbor} \footnote{\url{https://github.com/innolitics/natural-neighbor-interpolation}} that implements a fast Discrete Sibson interpolation \citep{interpolation}. 

Weighted surface averages of an MHD quantity $\mathcal{Q}$ with respect to a quantity $\sigma$ are defined as
\begin{equation}
\langle \mathcal{Q} \rangle_\sigma := \frac{\int dA \: \sigma \: \mathcal{Q}}{\int dA \:  \sigma},
\end{equation}
where $dA:= d\theta d\phi \sqrt{-g}$. A time-average of a surface-average is defined as
\begin{equation}
\langle \langle \mathcal{Q} \rangle  \rangle := \frac{1}{\Delta t} \int^{t_{i}}_{t_{f}} dt \langle \mathcal{Q} \rangle,
\end{equation}
where we always sum over a given time interval after the initial transient. 

\section{Results}
\label{sec-results}

\subsection{Overview of the system and previous studies}

In the steady-state of an equal-mass SMBBH, a lump orbits the edge of the circumbinary disk at an average  frequency $\langle \Omega_{\rm lump} \rangle = 0.28 ~ \Omega_{\mathcal{B}}$ \citep{Noble12, Shi12, DOrazio2013, DOrazio2016, Farris15a, Farris15b, armengol2021, noble2021} modulating the accretion into the cavity. When one of the BHs passes near the lump, it peels off part of lump's inner edge, forming a stream that feeds the black hole with a beat frequency of $\Omega_{\rm beat} :=  \Omega_{\mathcal{B}} - \langle \Omega_{\rm lump} \rangle \sim 0.72 \: \Omega_{\mathcal{B}}$. This stream is almost ballistic, formed by fluid particles with relatively low angular momentum \citep{Shi2015}. This material can start orbiting the black hole as it approaches, forming a mini-disk. The maximum size of the mini-disk is determined by the tidal truncation radius of the binary, or Hill's sphere. The residence time of matter in the mini-disks is determined by the ratio of the truncation radius and the radius of the ISCO. At close relativistic separations, such as the ones here, the mini-disks will be out of inflow equilibrium with the circumbinary lump accretion, and thus the masses oscillate quasi periodically in a filling and depletion cycle \citep{Bowen2018, Bowen2019}.

For relativistic binaries, the tidal truncation radius is approximately at $r_t \sim 0.4 \: r_{12}(t)$ \citep{Bowen17}, similar to the Newtonian value, estimated to be $\sim 0.3 \: r_{12}$ \citep{Paczynski:1977, Papaloizou:1977a, ArtymLubow94}. Since spin is a second order effect in the effective potential \citep{armengol2021}, mild values of spin do not change the truncation radius significantly for binary separations greater or close to $20M$. The most relevant difference between our two simulations, \ssim{} and \nossim{}, is the location of the ISCO: $r_{\rm ISCO}(\chi=0.6) = 2.82 \: M_{\rm BH_{i}} $ for \ssim{}, and $r_{\rm ISCO}(\chi=0.0)= 5.0 \: M_{\rm BH_{i}} $ for \nossim{} (both in given here in harmonic coordinates). The smaller ISCOs of the spinning black holes allow material with lower angular momentum to maintain circular orbits closer to the BH instead of plunging in directly (see Section \ref{sec-struct}). 

In the following sections, we analyze how the size of the ISCO plays a role in the accretion rate, inflow time, and periodicities, as well as in the structure of the mini-disks. We will also examine how the presence of an ergosphere in the spinning case helps the black holes to produce more Poynting flux. We will also analyze how the variability of the fluxes is connected with the variability of the accretion rate, dominated by the lump. 

\subsection{Mass evolution, accretion rate, and inflow time}
\label{sec-mdot}
\begin{figure}[htb!]
  \centering
  \includegraphics[width=\columnwidth]{\figfolder/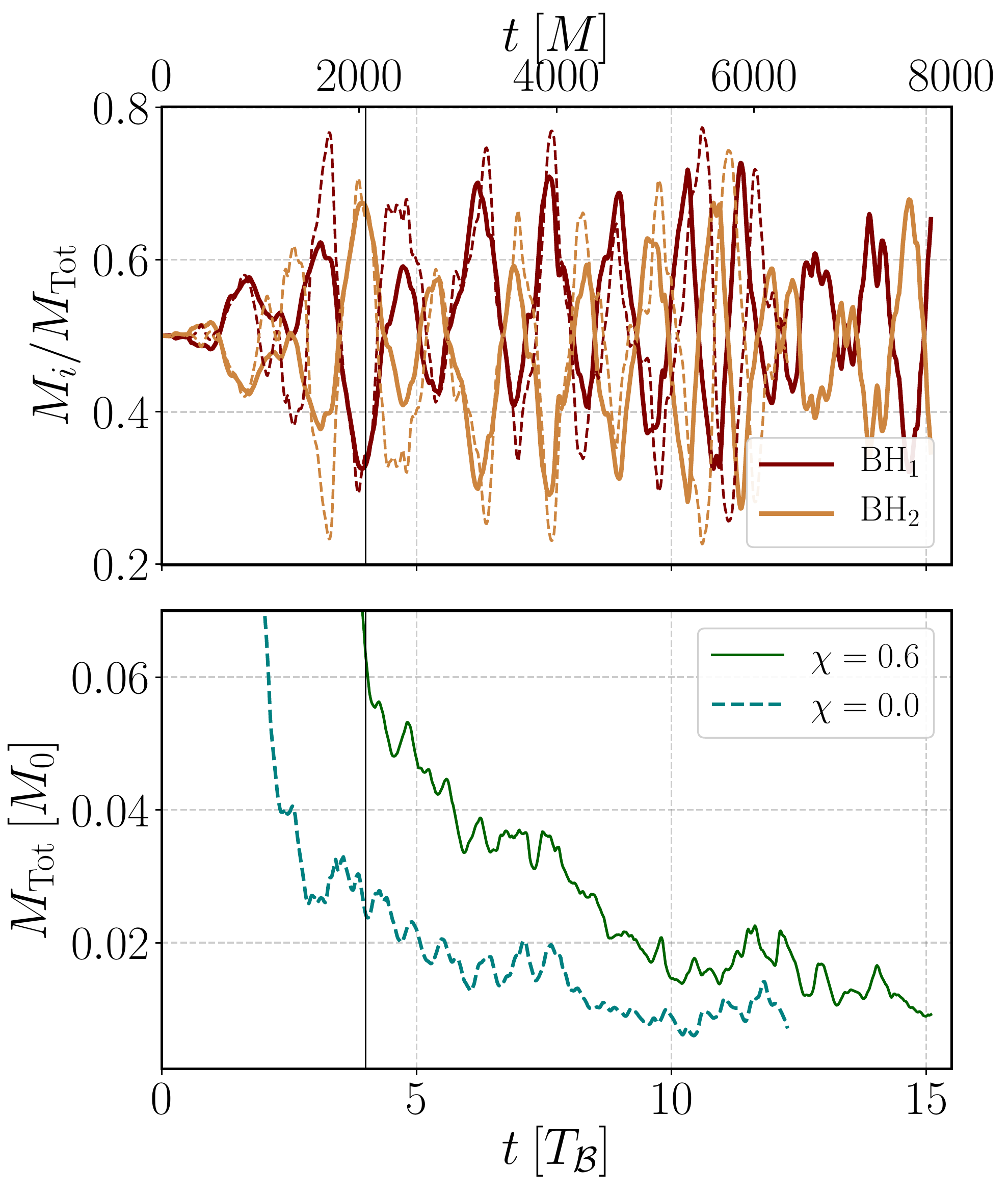}
  \caption{Upper panel: the mass fraction evolution $M_i$ of each minidisk for \ssim{} (solid lines) and \nossim{} (dashed lines), where we define $M_{\rm tot}:= M_1(t)+M_2(t)$. Lower panel: total mass evolution for \ssim{} (solid line) and \nossim{} (dashed line), where $M_0:= M_1(0)+M_2(0)$. The black line indicates the end of the transient phase.}
  \label{fig-massminidisk}
\end{figure}

We start the analysis by calculating the \textbf{integrated rest-mass} of each mini-disk, defined as
\begin{equation}
M := \int^{r_t(t)}_{r_{\mathcal{H}}}   dV \: \rho u^{t},
\label{eq-mass}
\end{equation}
where $r_t(t):= 0.4 \: r_{12}(t)$ is the truncation radius, $r_{\mathcal{H}}$ is the BH horizon, and $dV:= \sqrt{-g} d^3x$ is the volume element. In both simulations there is an initial transient due to the initial conditions that lasts approximately $\sim 3$ orbits for \nossim{} and $\sim 4$ orbits for \ssim{} (see Figure \ref{fig-massminidisk}). Both simulations start with two quasi-equilibrated mini-tori around the holes, with a specific angular momentum distribution adapted specifically for non-spinning black holes, following the prescription described in \cite{Bowen17}. Since we are also using this initial data for the spinning simulation, the initial tori have an excess of angular momentum, making the transient slightly longer in \ssim{}. We analyze each simulation after this transient, marked in the plots as a vertical line. As a time unit, we use $T_{\mathcal{B}} = 530 \: M$, the average binary period of the non-spinning simulation.

We find that although the mini-disks of \ssim{}, like those in \nossim{}, go through a filling-depletion cycle, the mini-disks around spinning BHs in \ssim{} are more massive than in \nossim{} by a factor of $2$ through most of the evolution (Figure~\ref{fig-massminidisk}), although they both follow the same decay. When the mass fraction of a mini-disk is more than $50\%$ of the total mass, we say that the disk is in its \textit{high state}; otherwise, it is in its \textit{low state}. The cycle of the mass fraction is similar in both simulations, although marginally smaller in amplitude for \ssim{}. The frequency of the cycle is associated with the orbital frequency and thus is higher in the spinning case, as can be seen plainly after $\sim 8$ orbits (upper panel {color{blue} of Fig.~\ref{fig-massminidisk}}). On the other hand, at $t=11 \: T_{\mathcal{B}}$, we observe a slight increase of mass in the system. Because the lump grows and oscillates radially around the cavity \citep{armengol2021}, it generates stronger accretion events onto the black holes with a lower frequency rate. 

We also analyze the accretion rate evolution at the horizon in the black hole rest-frame, which is given by
\begin{equation}
\dot{M} := \oint_{r_{\mathcal{H}}}  d\bar{A} \: u^{\bar{r}} \rho.
\label{eq:massflux}
\end{equation}

\begin{figure}
  \includegraphics[width=\columnwidth]{\figfolder/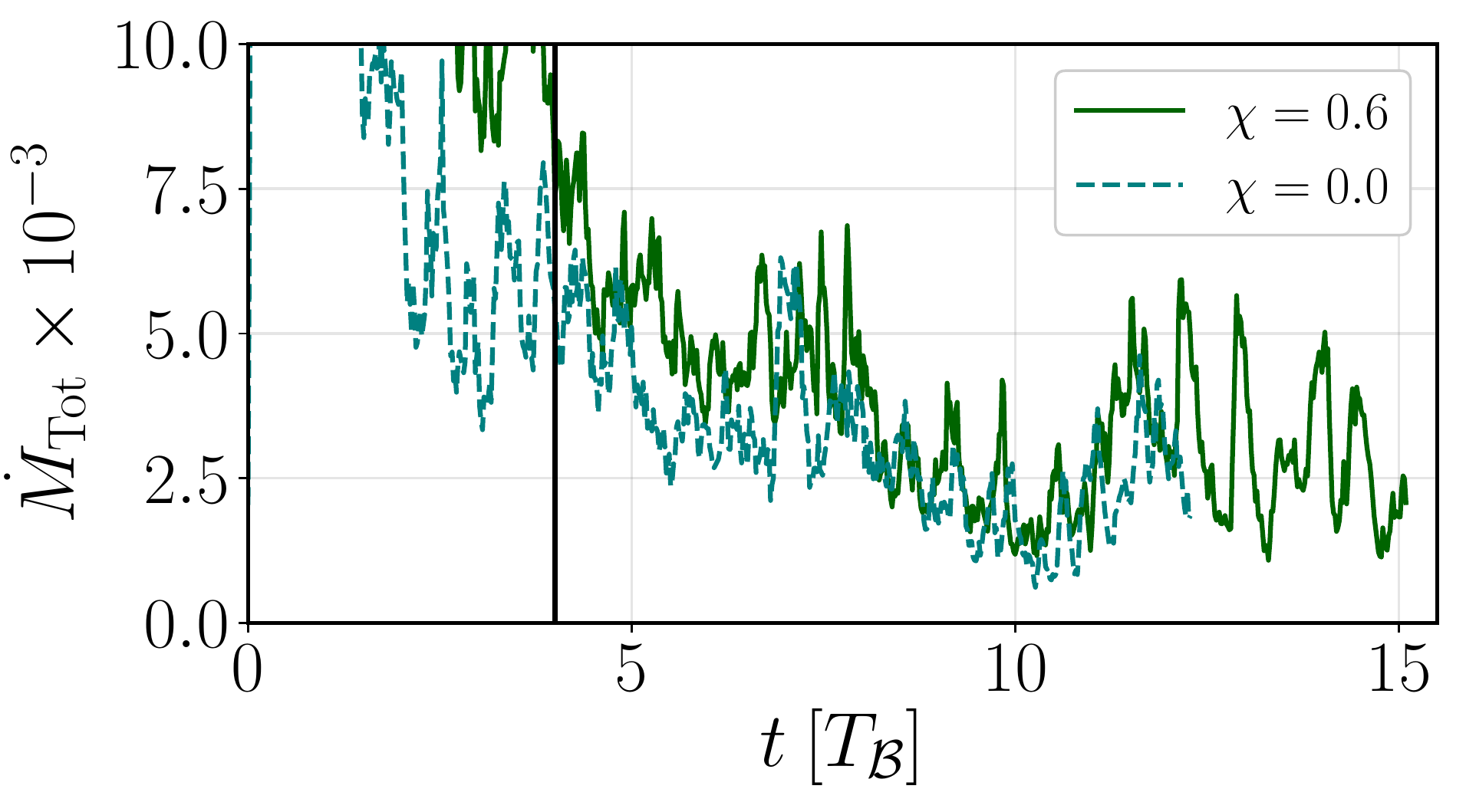}
  \caption{Total accretion rate evolution $\dot{M}_{\rm Tot} \equiv \dot{M}_{\rm BH_2} + \dot{M}_{\rm BH_2}$ in \ssim{} (solid lines) and \nossim (dashed lines)}
  \label{fig-accrate}
\end{figure}

\begin{figure}
  \includegraphics[width=\columnwidth]{\figfolder/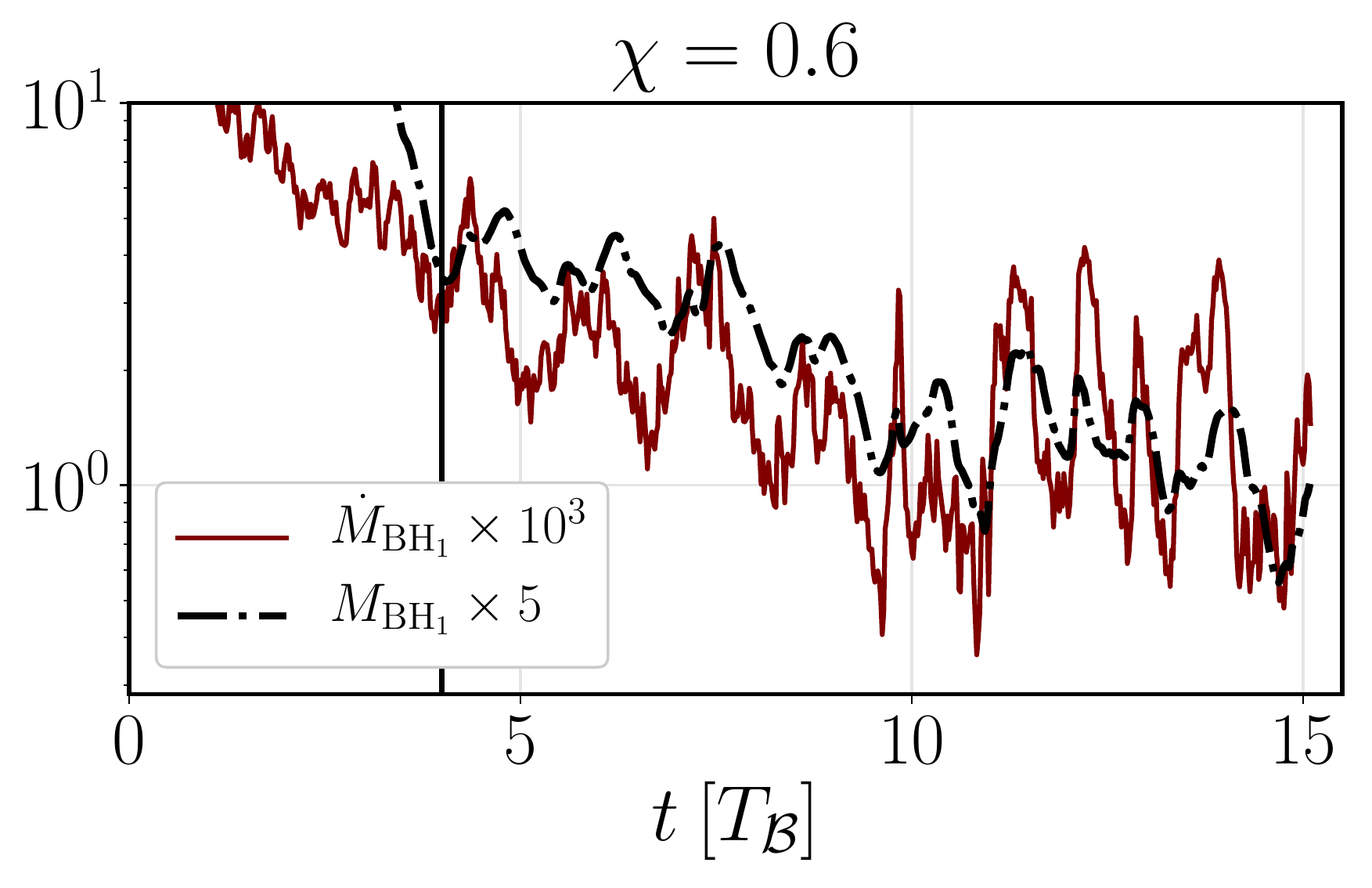}
  \caption{Accretion rate ($\dot{M}_{\rm BH_1}$) in red solid lines and mass ($M_{\rm BH_1}$) in black dot-dashed lines for a mini-disk around BH$_1$ in \ssim{}. Both quantities were rescaled for plotting. }
  \label{fig-accratemassbh1}
\end{figure}

Here, and in the remainder of this paper, overbars indicate (harmonic) coordinates whose origin is the center of one of the black holes. We plot the sum of the accretion rate in each BH, $\dot{M}_{\rm Tot}$, for each simulation in Figure \ref{fig-accrate}. We find that the accretion rate evolution is overall similar after the transient in both \ssim{} and \nossim{}. Moreover, the time-dependence of the mini-disk mass is, to a first approximation, a smoothed version of the accretion rate's time-dependence, e.g. see Figure \ref{fig-accratemassbh1} for BH$_1$ in \ssim{}.

\begin{figure}[htb]
  \includegraphics[width=\columnwidth]{\figfolder/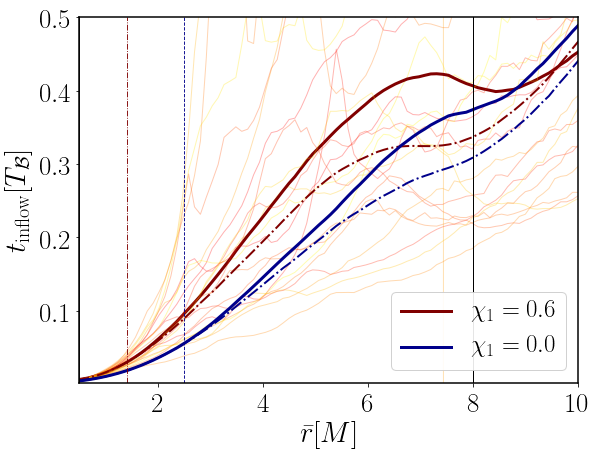}
  \caption{Inflow time as a function of radius  in harmonic coordinates for BH$_1$ in \ssim{} (maroon) and \nossim{} (dark blue). Solid lines are time averages over the first half of the simulation, while dot-dashed lines are time averages of the last half of the simulation. Thin lines show the instantaneous inflow time every $300M$ for the spinning case. The maroon (dark blue) vertical line represents the ISCO of the spinning (non-spinning) black hole. The vertical black line is the initial truncation radius of the binary.}
  \label{fig-inflow}
\end{figure}

The differences in the masses per cycle are closely related to the inflow time of particles in the mini-disk. It is useful then to define an (Eulerian) inflow time as the characteristic time for a fluid element to move past a fixed radius $r$ \citep{KHH05}. On average, this can be defined as
\begin{equation}
t_{\rm inflow}^{-1} :=  \frac{1}{\bar{r}} \Big \langle \langle V^{\bar{r}} \rangle_{\rho} \Big \rangle,
\end{equation}
where $V^{\bar{r}}:= u^{\bar{r}}/u^{\bar{t}}$ is the transport velocity. In Figure \ref{fig-inflow}, we show the inflow time as a function of coordinate radius for BH$_1$, averaged in time for the first part (solid lines) and second part (dashed lines) of both simulations. Inside the truncation radius, the inflow time is consistently longer in \ssim{} than in \nossim, generally by tens of percent. 

As the binary shrinks, the inflow time at the truncation radius diminishes from $\sim 0.41 \: T_{\mathcal{B}}$ to $\sim 0.31 \: T_{\mathcal{B}}$ for the spinning simulation. The average inflow time at the truncation radius is well below the beat period, $ T_{\rm beat}  = 1.3 T_{\mathcal{B}}$, on which the mini-disk refills. Our mean inflow time is also significantly shorter than typical inflow timescales of Keplerian orbits around single black holes \citep{KHH05}. This suggests that accretion in our mini-disks is driven by different mechanisms than single BH disks. Moreover, the similar measures of accretion rates at the horizon for \ssim{} and \nossim{} (see again Figure \ref{fig-accrate}), seem to imply a shared accretion mechanism between spinning and non-spinning systems (see Sec.~\ref{sec-struct}).

\begin{figure}[ht!]
  \includegraphics[width=\columnwidth]{\figfolder/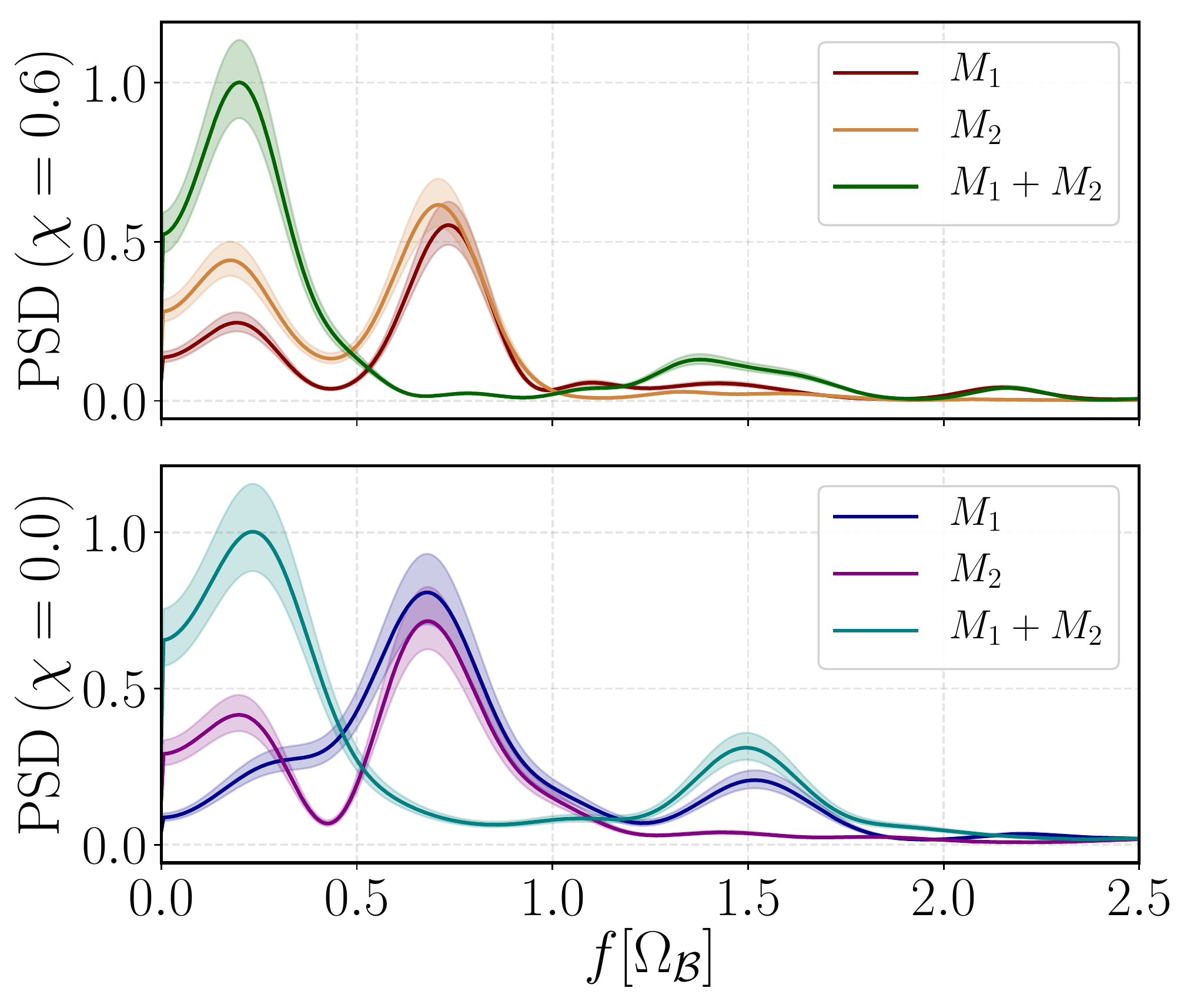}
  \caption{Power spectral density of the mini-disk's masses for \ssim{} (upper panel) and \nossim{} (lower pannel) using a Welch algorithm with a Hamming window size and a frequency of $10M$. The confidence intervals at $3 \sigma$ are shown as shadowed areas.}
  \label{fig-psd}
\end{figure}

The quasi-periodic behavior of the system can be described through a power density spectrum (PSD) of the mini-disk's masses. In Figure \ref{fig-psd} we show the PSD, normalized with the power of the highest peak within the simulation. Most features of the system's quasi-periodicities are shared in both simulations and were described in detail in \cite{Bowen2019}. We find, still, some interesting differences. The PSDs for $M_1$ and $M_2$ taken individually peak at the beat frequency in both simulations. The PSD of the total mass of the mini-disks, $M_1+M_2$, has a peak at $2\Omega_{\rm beat}$ in \nossim{}, while the latter is severely damped in \ssim{}. Indeed, if the individual masses vary with a characteristic frequency $\Omega_{\rm beat}$, and these are out of phase with the same amplitude, we expect their sum to vary with $2\Omega_{\rm beat}$. In \ssim{}, however, the inflow time of the mini-disks is larger and the depletion period of a mini-disk briefly coexists with the filling period of the other mini-disk, reducing the variability of the total mass. On the other hand, the beat frequency is slightly higher for \ssim{} ($\Omega_{\rm beat} = 0.71 \Omega_{\mathcal{B}} $) than \nossim{} ($\Omega_{\rm beat}= 0.68 \Omega_{\mathcal{B}} $) as the orbital frequency of the spinning BHs is higher. Further, since we evolved the binary for longer, in \ssim{} we find a more prominent amplitude at $\sim 0.20 \Omega_{\mathcal{B}}$. We can associate this low-frequency power to the radial oscillations of the lump around the cavity that produces additional accretion events \citep{armengol2021}. We observe this frequency is closely related but different from the orbital frequency of the lump at $0.28 \Omega_{\mathcal{B}}$, which could be related to the orbital frequency increasing during the inspiral (notice we measure the PSD at fixed frequencies).

Finally, because we use a spherical grid with a central cutout, we cannot analyze the effects of the sloshing of matter between mini-disks \citep{Bowen17}. To estimate how much mass we lose through the cutout, we compute the accretion rate at the inner boundary of the grid. This mass loss constitutes only $5 \%$ of the total mass accreted by the BHs throughout the simulation, although the instantaneous accretion can be close to $20\%$ of the accretion onto a single BH.  We do not expect this small mass loss to alter the main conclusions of this work, namely, the differences between mini-disks in spinning and non-spinning BBH.
 
\subsection{Structure and orbital motion in mini-disks}
\label{sec-struct}

In this section, we analyze in detail how the structure of the mini-disks in \ssim{} compares with mini-disks in \nossim{}. We first focus on how the spin changes the surface density distribution and the azimuthal density modes in the mini-disks. We then investigate the angular momentum of the fluid and how it compares with the angular momentum at the ISCO. 

\begin{figure}[ht]
  \includegraphics[width=\columnwidth]{\figfolder/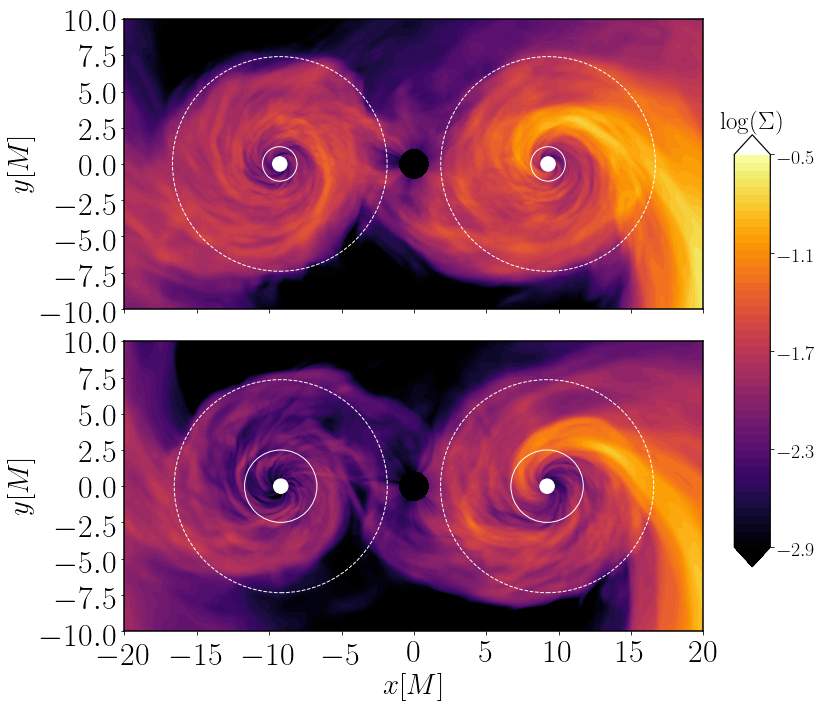}
  \caption{Surface density snapshot for \ssim{} (upper row) and \nossim{} (lower row) at $t=4000M$ and $t=4060M$ respectively, where the phase of the binary is the same in both simulations. White dashed lines indicate the truncation radius and solid white lines indicate the ISCO. The sense of rotation of the binary is counter-clockwise.}
  \label{fig-surfden-2D}
\end{figure}

\begin{figure}[ht]
  \includegraphics[width=\columnwidth]{\figfolder/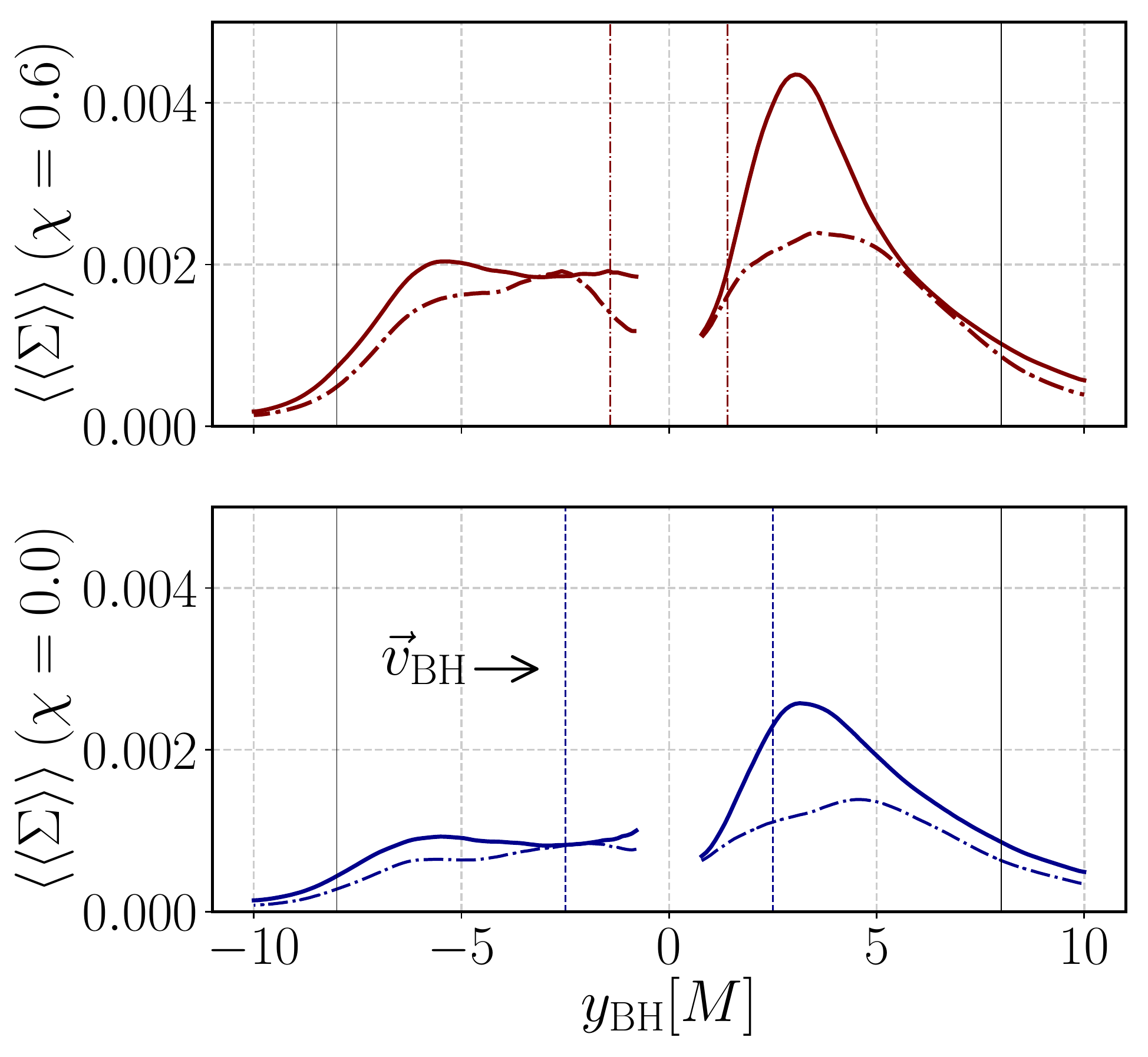}
  \caption{Surface density average in the azimuthal ranges $\Delta \phi_1=(\pi/4,3\pi/4) $ (positive $y_{\rm BH}$-axis) and $\Delta \phi_2=(5\pi/4,7\pi/4)$ (negative $y_{\rm BH}$-axis) for BH$_1$ in \ssim (upper panel) and \nossim (lower panel). Solid lines represent a time average on the high state of the cycle, while dot-dashed lines represent a time average over the low state. For reference, we indicate the direction of the orbital BH velocity. Dashed vertical lines indicate the position of the ISCO.}
  \label{fig-surfden}
\end{figure}

The surface density is defined as $\Sigma(t, r,\phi) := \int  d\theta \rho \sqrt{-g}/\sqrt{-\sigma}$, where \firstrev{we use $\sqrt{-\sigma}= \sqrt{g_{\phi \phi} g_{rr}}$ as the surface metric of the equatorial plane.} In Figure \ref{fig-surfden-2D}, we plot the surface density in both \ssim{} and \nossim{}, for the same orbital phase at the $7$th orbit. In this plot, the mini-disk around BH$_1$ (right side) is in the peak of the mass cycle. In both simulations we can clearly notice the lump stream plunging directly into the hole. There is also circularized gas orbiting BH$_1$ in both simulations, but there is much more of it in \ssim{}. On the other hand, we observe that BH$_2$ (left side), in its low state, has a noticeable disk structure in \ssim{}, while the material is already depleted for \nossim{}.

We can quantify these differences in structure by computing the average surface density over two ranges of $\bar{\phi}$, as measured in the BH frame, representing the front and back of the mini-disk with respect to the orbital motion. We define
\begin{equation}
\langle \Sigma(r,t) \rangle := \frac{ \int_{\Delta \bar{\phi}} d\bar{l} \: \Sigma }{\int_{\Delta \bar{\phi}} d\bar{l} },
\end{equation}
where $d\bar{l}:= d\bar{\phi} \sqrt{g_{\bar{\phi} \bar{\phi}} (\bar{\theta}=\pi/2)}$.  In Figure \ref{fig-surfden}, we plot  $\langle \langle \Sigma(r) \rangle \rangle $ for $\Delta \bar{\phi}_1=(\pi/4,3\pi/4) $ and $\Delta \bar{\phi}_2=(5\pi/4,7\pi/4)$, averaging in time over the high and low state of the mini-disk separately. In the high-state, the mini-disk accumulates more material at the front while the back of the mini-disk is flatter. In the low-state, both simulations show a flatter profile, with a slightly higher density at the front. The asymmetry between front and back arises because of the orbital motion of the black holes, capturing and accumulating the stream material as they orbit. In \ssim{}, the density profile is steeper near the ISCO for high and low states. In the outer part of the mini-disk, the slope of the surface density for both \nossim{} and \ssim{} have a similar profile, indicating a common truncation radius. The density is higher in \ssim{} by a factor of $\sim 2$ in both states.

\begin{figure}[htb]
  \includegraphics[width=\columnwidth]{\figfolder/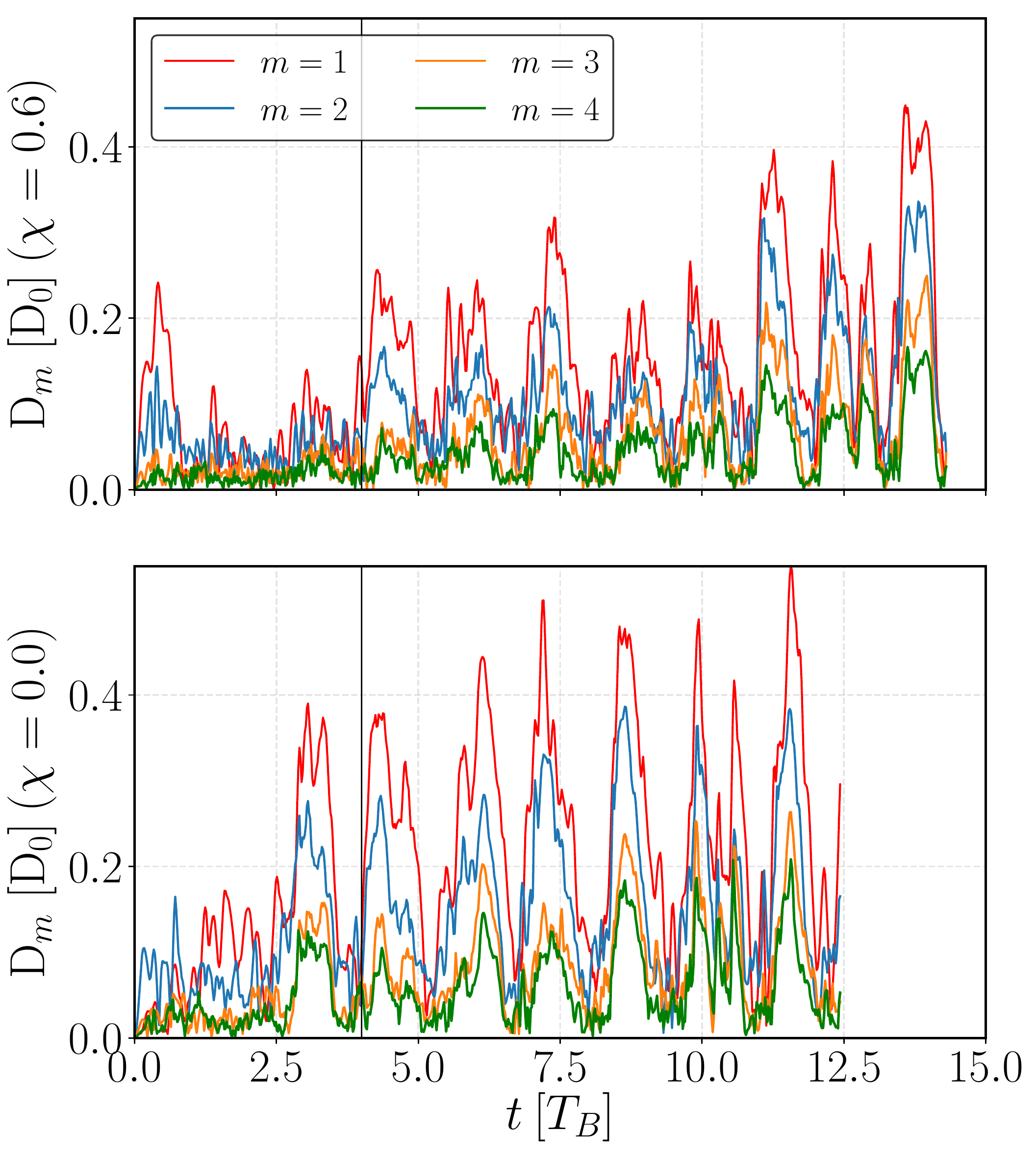}
  \caption{Azimuthal density modes for BH$_1$ in \ssim{} (upper panel) and \nossim{} (lower panel) normalized with the zero mode $D_0$. The black vertical line represents the end of the transient in \ssim.}
  \label{fig-modes}
\end{figure}

Another important property of mini-disks in relativistic binaries is the presence of non-trivial azimuthal density modes. When the accreting stream of the lump impacts the mini-disk, it generates a pressure wave, forming strong spiral shocks \citep{Bowen2018}. This induces an $m=1$ density mode in the mini-disk that competes with the $m=2$ mode excited by the tidal interaction of the companion black hole. The spiral wave patterns can be analyzed decomposing the mini-disk rest-mass  density in azimuthal Fourier modes \citep{Zurek1986}, $\rho(\bar{\phi})\equiv \sum_m D_{m} \exp{(-im\bar{\phi})}$, where
\begin{equation}
D_{m} := \int^{r_t(t)}_{r_{\mathcal{H}}} d\bar{V} \rho \exp{(-im\bar{\phi})}.
\end{equation}

\begin{figure}[htb]
  \includegraphics[width=\columnwidth]{\figfolder/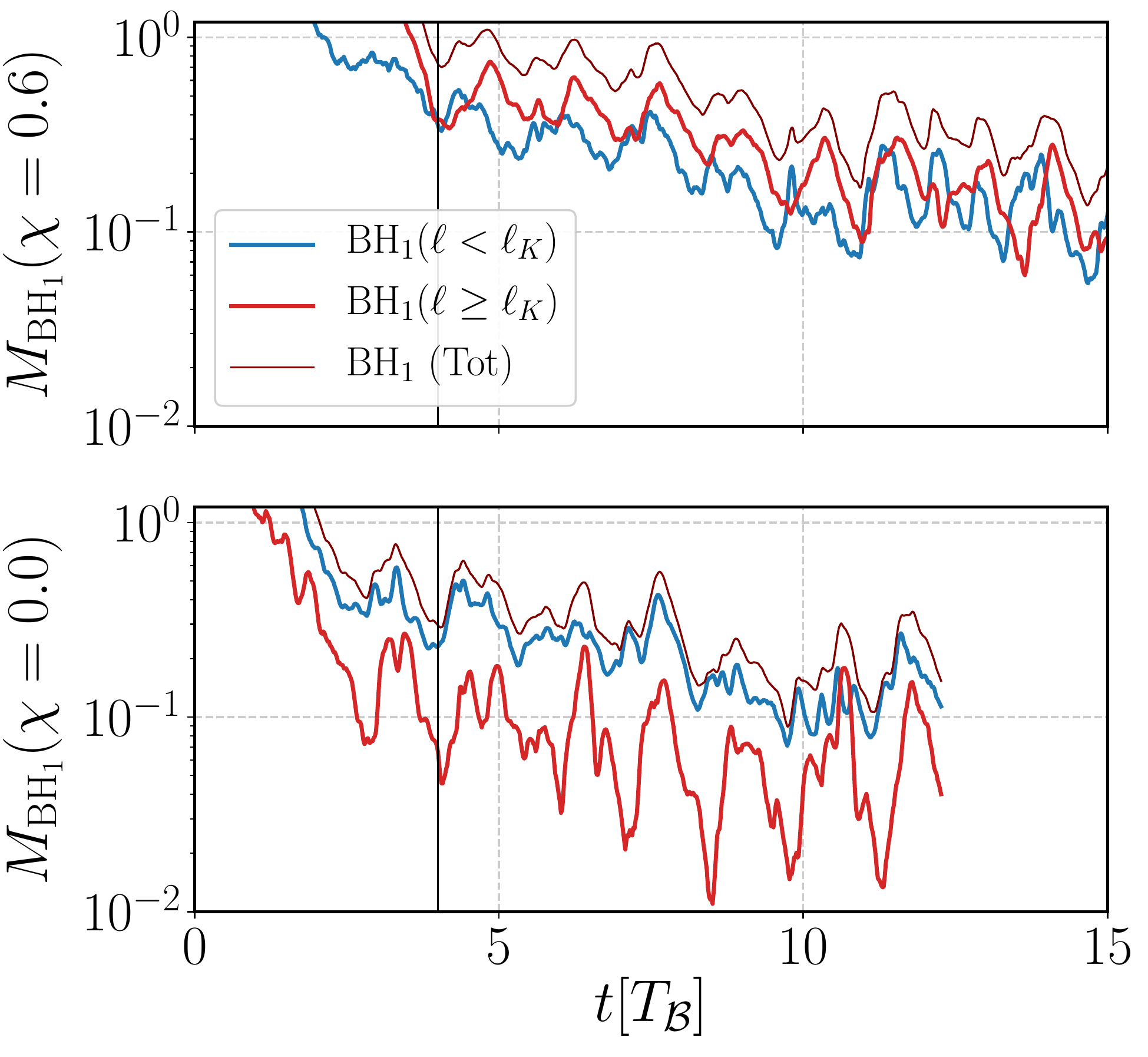}
  \caption{Sub-Keplerian and (super-)Keplerian components of the mass for BH$_1$ in \ssim{} (upper panel) and \nossim{} (lower panel)}
  \label{fig-circmass}
\end{figure}

Let us compare these modes in \nossim{} and \ssim{} for BH$_1$. From Figure \ref{fig-modes}, we observe that both simulations share common features. In both simulations, the mini-disks are mainly dominated by $m=1$ modes, followed closely by $m=2$ modes. We also observe important $m=3,4$ contributions. The modes are excited in the high state of the cycle, where the mini-disks increase their mass and the stream is accreted onto the black holes. In \nossim{}, the amplitudes of the modes are noticeably larger than \ssim{}, while in the latter the amplitudes grow as the system evolves. The growth of modes in \ssim{} is correlated with the mass decrease of the mini-disks. This behavior could indicate that, as the mini-disks become less massive, the density modes are really representing the single-arm stream of the lump that plunges directly into the hole. Azimuthal modes grow to large amplitudes in the high phase of a mini-disk only when the material orbiting the BH is less or equally massive than the material with low angular momentum that plunges from the lump, occurring around $10 T_{\mathcal{B}}$ in \ssim{} (see Figure \ref{fig-circmass} and discussion below). This is another consequence of the disk-like structure of the mini-disks surviving for longer time in the spinning case.

\begin{figure}[htb]
  \includegraphics[width=\columnwidth]{\figfolder/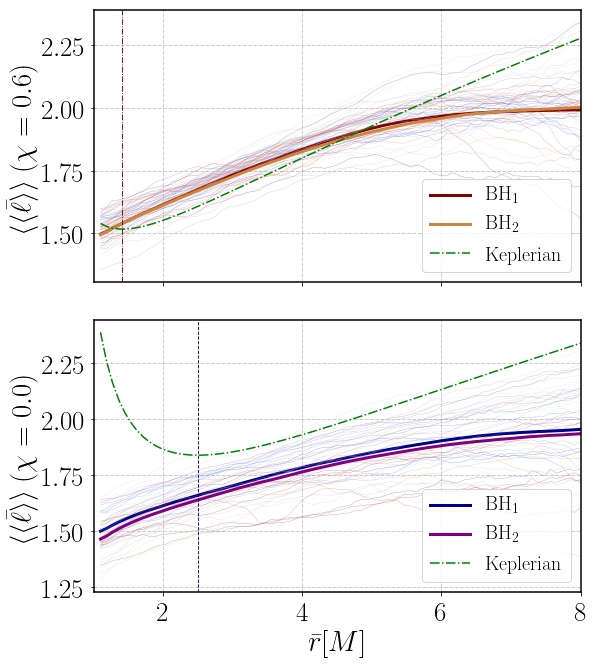}
  \caption{Specific angular momentum as a function of radius for \ssim{} (upper panel) and \nossim{} (lower panel) for both BHs. The time-averages are in solid lines, and the individual values are the very thin lines. The Keplerian value is plotted in dashed green lines.}
  \label{fig-samvsr}
\end{figure}

Our analysis so far indicates that a considerable amount of the matter in the mini-disk region is plunging directly from the lump to the black hole. In order to further analyze the orbital motion of the fluid in the mini-disk, we compute the density-weighted specific angular momentum in the BH frame:
\begin{equation}
\langle \bar{\ell} \rangle_{\rho} :=  \langle -u_{\bar{\phi}}/u_{\bar{t}} \rangle_{\rho}.
\end{equation}

In Figure \ref{fig-samvsr}, we show the time-average of $\langle \bar{\ell} \rangle_{\rho}$ for both BHs and both simulations. In absolute terms, $\langle \bar{\ell} \rangle_\rho$ is nearly the same for both the spinning and non-spinning cases, with the spinning case only slightly greater.  This is because the specific angular momentum of the material that falls into the cavity is essentially determined by the stresses at the inner edge of the circumbinary disk. These stresses are determined by binary torques and the plasma Reynolds and magnetic stresses \citep{Shi12,Noble12}. Indeed, in \cite{armengol2021} we found that these quantities depend weakly on spin outside the cavity.
On the other hand, their relation to their respective Keplerian (circular orbit) values, $\bar{\ell}_K(\bar{r},\chi)$, is quite different because they depend strongly on the spin. In \ssim{}, the distribution of the angular momentum tracks closely the Keplerian value. For \nossim{}, the behavior is always sub-Keplerian on average. 

Let us assume that the angular momentum distribution of the circumbinary streams is independent of spin for a fixed binary separation and mass-ratio. In that case, our simulation data indicate that the angular momentum with which the streams arrive at the mini-disk would be greater than the ISCO angular momentum when the BH spin is $\chi> 0.45$. This estimate could serve as a crude criterion for determining whether mini-disks form in relativistic binaries.

We can also use the specific angular momentum to distinguish the material in the mini-disk with high angular momentum that manages to orbit the black hole from the low angular momentum part that plunges in. To do so, we recompute the mass as in equation~\eqref{eq-mass}, taking fluid elements with $\bar{\ell}< \bar{\ell}_K$ and $\bar{\ell} \geq \bar{\ell}_K$ separately. In Figure \ref{fig-circmass} we plot the evolution of the sub-Keplerian and super-Keplerian mass components for BH$_1$ in \ssim{} and \nossim{}. In \ssim{} after the initial transient, a little more than half of the mass comes from relatively high angular momentum fluid. As the system inspirals, however, the truncation radius decreases, and the masses of these two components become nearly equal. In \nossim{}, on the other hand, most of the fluid has relatively low angular momentum. This sub-Keplerian component has roughly the same mass in \ssim{} and \nossim{}, while the mass of the high angular momentum component of the fluid is much greater in \ssim{}, as expected.

\begin{figure}[htb]
  \includegraphics[width=\columnwidth]{\figfolder/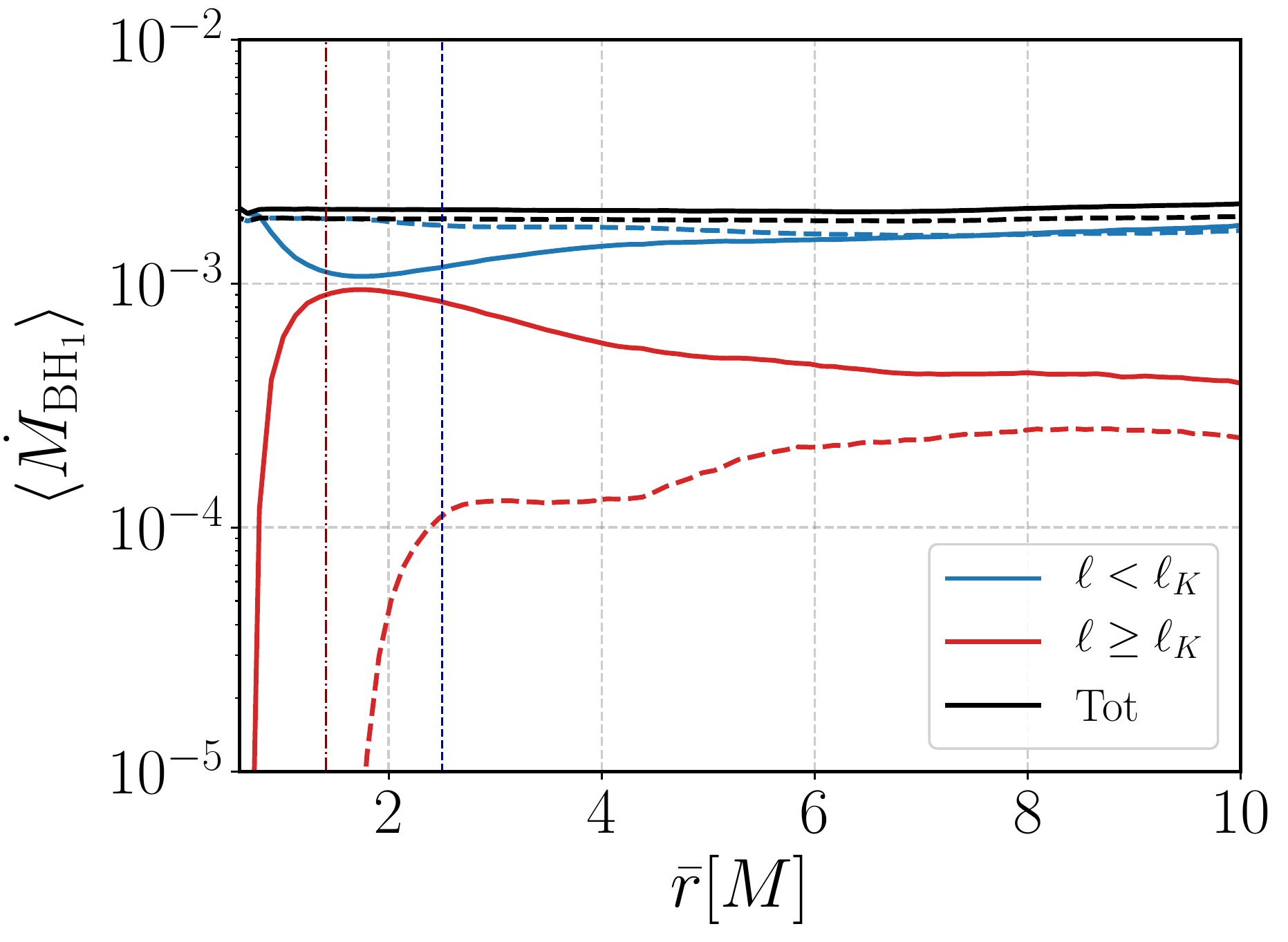}
  \caption{Time averaged accretion rate in BH$_1$ for \ssim{} (solid lines) and \nossim{} (dashed lines) considering particles with low (blue) and high (red) angular momentum. The vertical dashed blue and dot-dashed red lines mark the ISCO for \nossim{} and \ssim{}, respectively.}
  \label{fig-mdot-am}
\end{figure}

Although a fair amount of the mass in the mini-disk has relatively high angular momentum and manages to orbit the black hole in \ssim, the accreted mass onto the BH, in both simulations, is always dominated by the low angular momentum part that plunges directly. To demonstrate this, we compute the average accretion rates for low and high angular momentum particles as we did with the mass. Figure \ref{fig-mdot-am} shows that the total accretion rate onto the BH has a flat radial profile in both \ssim{} and \nossim{}, with very similar average values. Accretion by low angular momentum particles dominates at all radii, although the high angular momentum contribution becomes comparable to the low angular momentum one near the ISCO for \ssim{}.

\begin{figure}[htb]
  \includegraphics[width=\columnwidth]{\figfolder/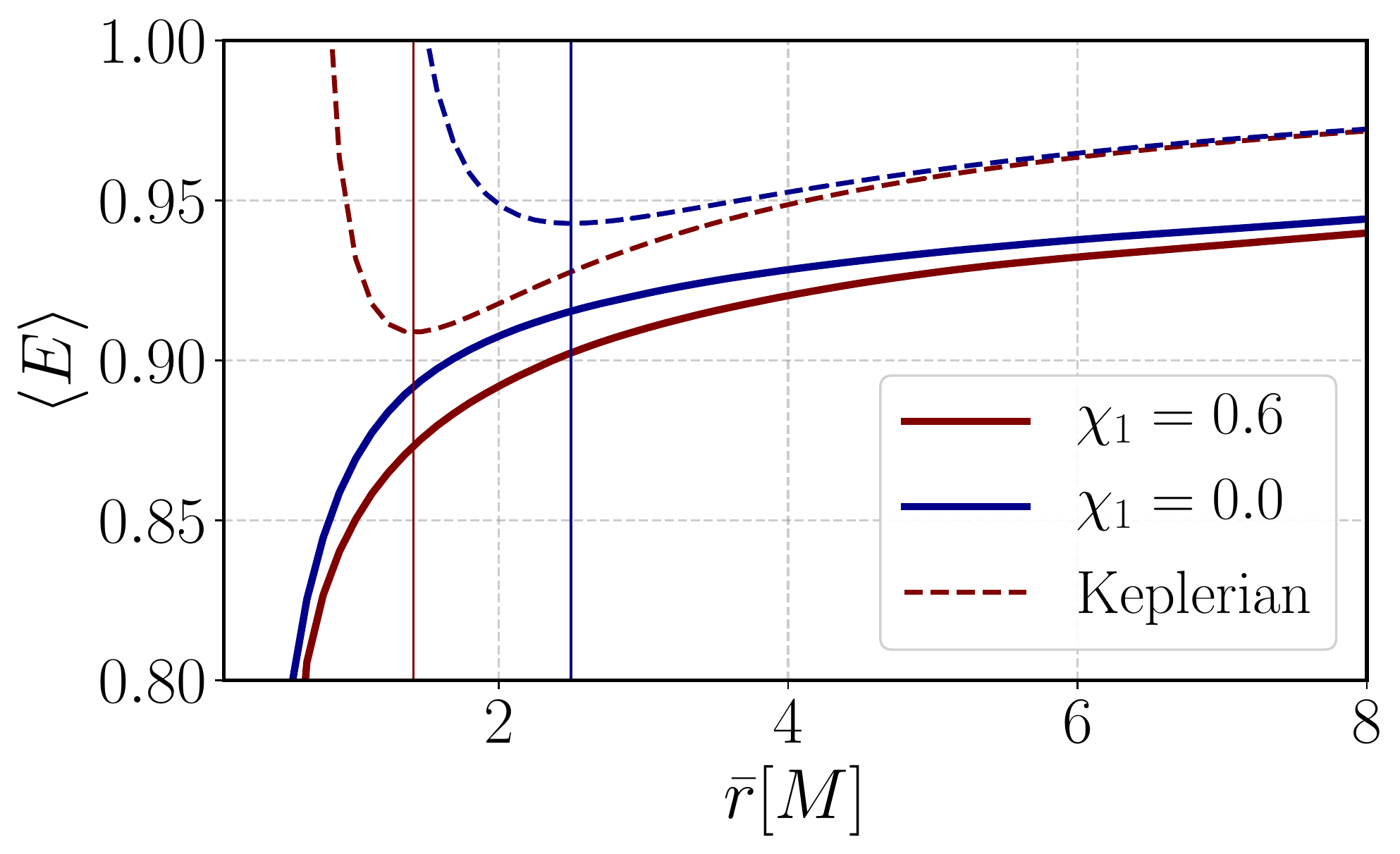}
  \caption{Density-weighted specific energy $E = - u_{\bar{t}}$, averaged in time as a function of radius for BH$_1$ in \ssim{} and \nossim{}. The dashed line represents the specific energy for a geodesic particle in circular motion in each case. Vertical lines indicate the location of the ISCO.}
  \label{fig-ut}
\end{figure}

We can also compute the density-weighted specific energy, $E:=\langle -u_{\bar{t}} \rangle_{\rho}$, the mass-weighted sum of rest-mass, kinetic, and binding energy for individual fluid elements. As can be seen in Figure~\ref{fig-ut}, on average, fluid in the mini-disks around the spinning black holes is more bound than in the non-spinning case. On the other hand, fluid in both \ssim{} and \nossim{} is more bound than particles on circular orbits. Near the ISCO, the specific energy drops sharply inward in both cases, as is often found when accretion physics is treated in MHD: stress does not cease at the ISCO when magnetic fields are present.

When the mini-disk is in its high state, the spiral shocks heat up the mini-disks and increase their aspect ratio, $h/r$. Given our cooling prescription, the entropy is kept close to its original value, regulating the aspect ratio to $h/r=0.1$. However, the gas scale height increases dramatically where the lump stream impacts the mini-disk. At the peak of the accretion cycle, the aspect ratio rises to $h/r \sim 2$, but the gas cools before the next accretion event.

\subsection{Electromagnetic and hydrodynamical fluxes}
\label{sec-emfluxes}

In this section we analyze the extraction of energy from the system in two forms: outward electromagnetic luminosities, arising from a Poynting flux, and unbound material. In particular, we analyze how these energy fluxes change with spin and how their variability is characterized by the same periodicity as the accretion. 

\begin{figure}[htb]
  \includegraphics[width=\columnwidth]{\figfolder/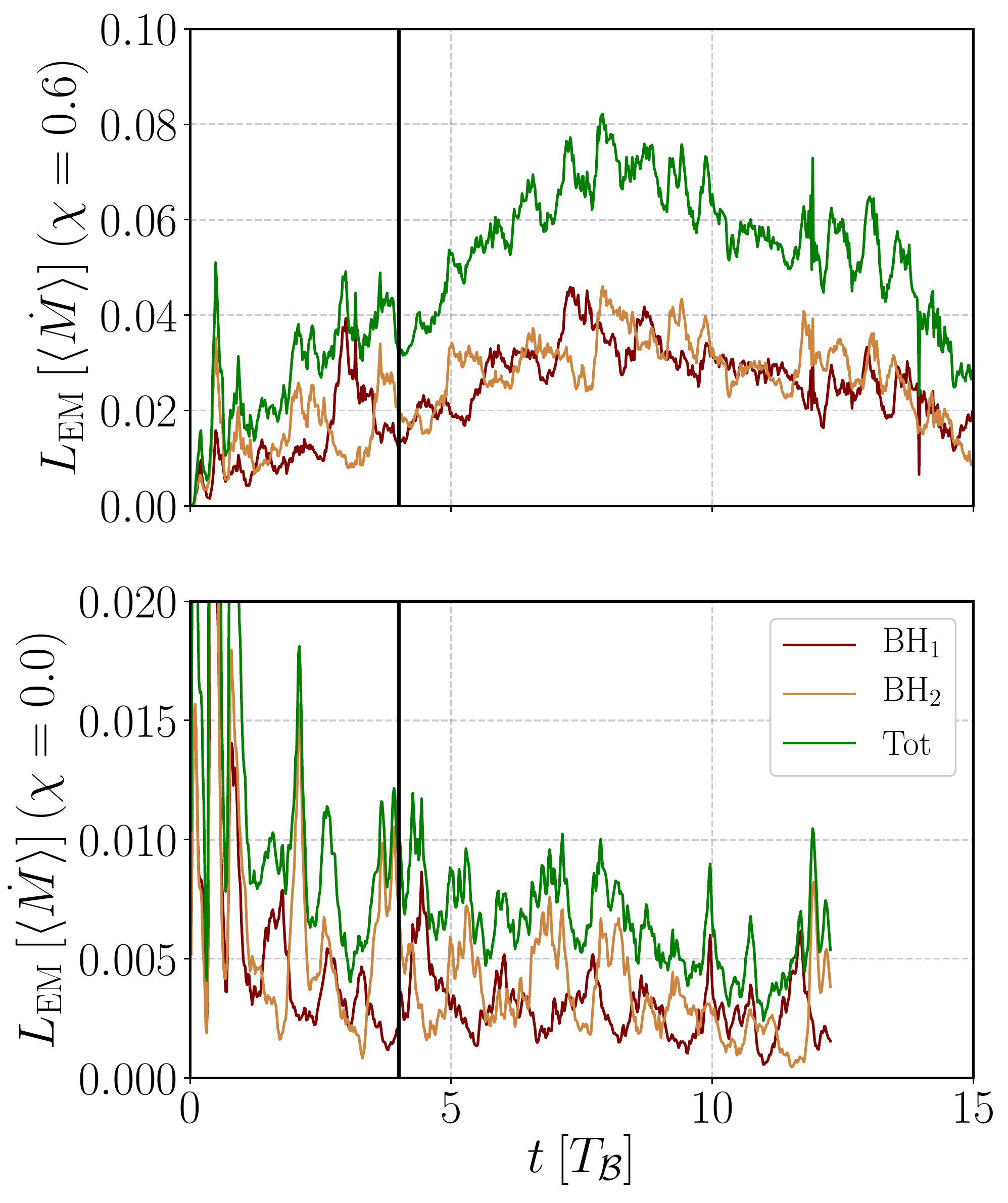}
  \caption{EM luminosity evolution in the BH frame from a sphere at $r=10M$ for \ssim{} (upper panel) and \nossim{} (lower panel).}
  \label{fig-poyntvst}
\end{figure}

The electromagnetic luminosity from each mini-disk  evaluated in the BH frame is:
\begin{equation}
L_{\rm EM} (t,\bar{r}) = \oint_{\bar{r}} d\bar{A} \: \mathcal{S}^{\bar{r}},
\end{equation}
where the Poynting flux is $\mathcal{S}^{\bar{i}}:=(T_{\rm EM})_{\: \: \bar{t}}^{\bar{i}}$. In Figure~\ref{fig-poyntvst}, we plot the EM luminosity as a function of time for \ssim{} and \nossim{}, evaluated on spheres of radius $\bar{r}=10 \: M $ that follow each black hole.  These luminosities are normalized to the average accretion rate $\langle \dot{M} \rangle = 0.002$, so they are equivalent to the rest-mass efficiency of the jets.  The most noteworthy element in Figure~\ref{fig-poyntvst} is that the EM luminosity is an order of magnitude larger in S06 than S0. In \nossim{}, there are no clear long-term trends while in \ssim{}, there is a secular growth of $L_{\rm EM}$ until $7.5  T_{\mathcal{B}}$, when it starts declining.

\begin{figure}[htb]
  \includegraphics[width=\columnwidth]{\figfolder/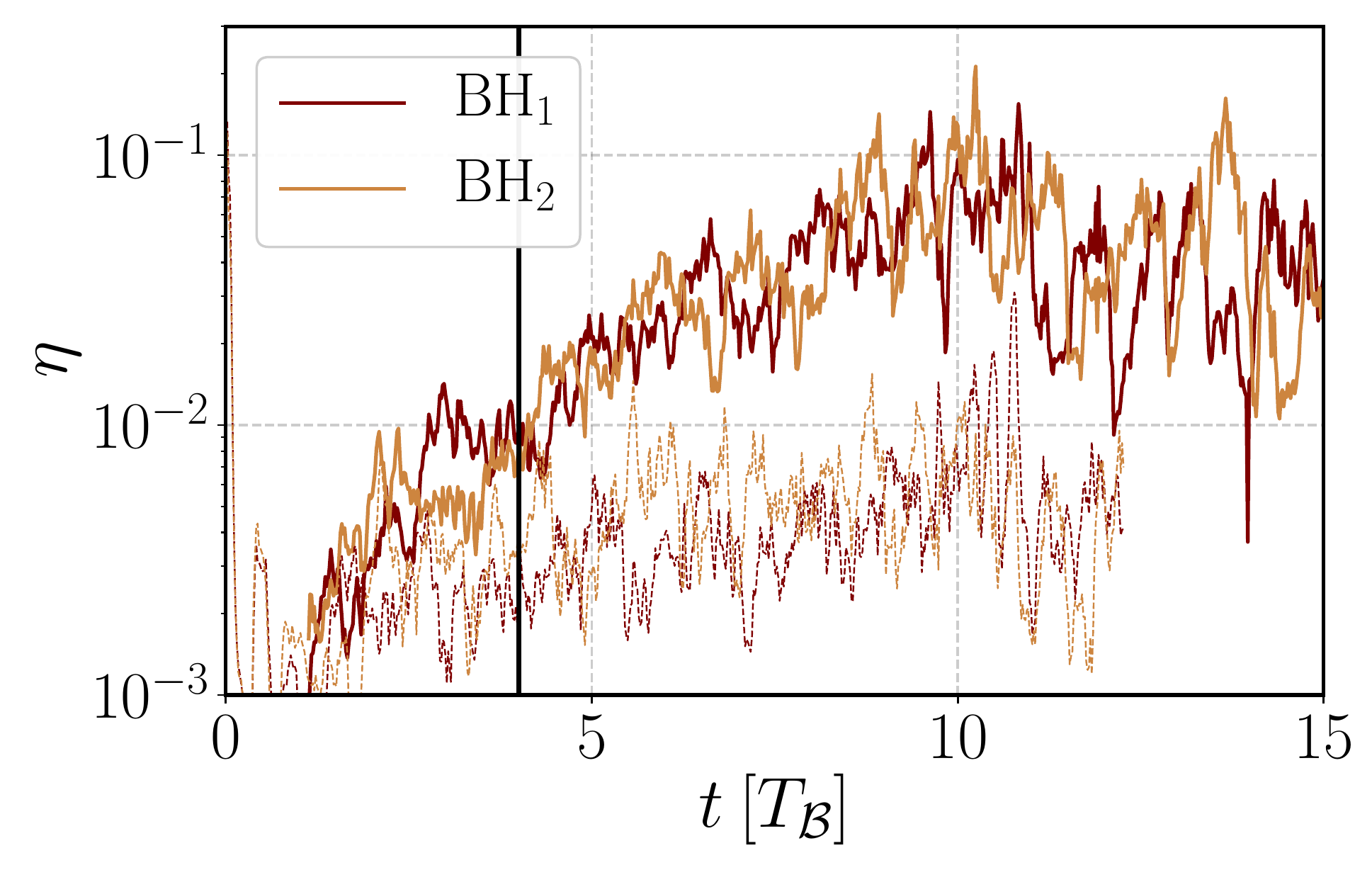}
  \caption{Instantaneous efficiency $\eta$ of the Poynting flux for \ssim{} (solid lines) and \nossim{} (dashed lines) in the BH frame measured at $\bar{r}=10M$.}
  \label{fig-efficiency}
\end{figure}

The instantaneous efficiency  $\eta = L_{\rm EM}(t)/\dot{M}(t)$ increases during the first several orbits, saturating at $\approx 0.05$ in the case of \ssim{}, but an order of magnitude lower for \nossim{},  as can be seen in Figure \ref{fig-efficiency}. The efficiency in \ssim{} with spin parameter 0.6 is rather larger than $0.013$, the value found by \cite{dVHKH05} for the efficiency of a single BH with specific angular momentum of 0.5 surrounded by a statistically time-steady disk. \secondrev{Both \nossim{} and \ssim{} present a similar secular growth of $\eta(t)$ until $10 T_{\mathcal{B}}$, where they slightly drop and plateau.}  \firstrev{Measuring fluxes in the comoving frame, we found that non-spinning black holes produce negligible EM luminosity, which is consistent with the fundamental idea of the Blandford-Znajek mechanism \citep{blanford1977, komissarov2001} and has been confirmed in many simulations \citep{McKG04,dVHKH05,HK06,TchekhovskoyChap}. However, because the BHs are orbiting around the center of mass this could power additional EM fluxes \citep{neilsen2011,palenzuela2010dual}. In order to capture the electromagnetic fluxes from the entire binary, we move to the center of mass frame and calculate the Poynting flux through a sphere of radius $r=100 \: M$ surrounding the binary system.}

\begin{figure}[htb]
  \includegraphics[width=\columnwidth]{\figfolder/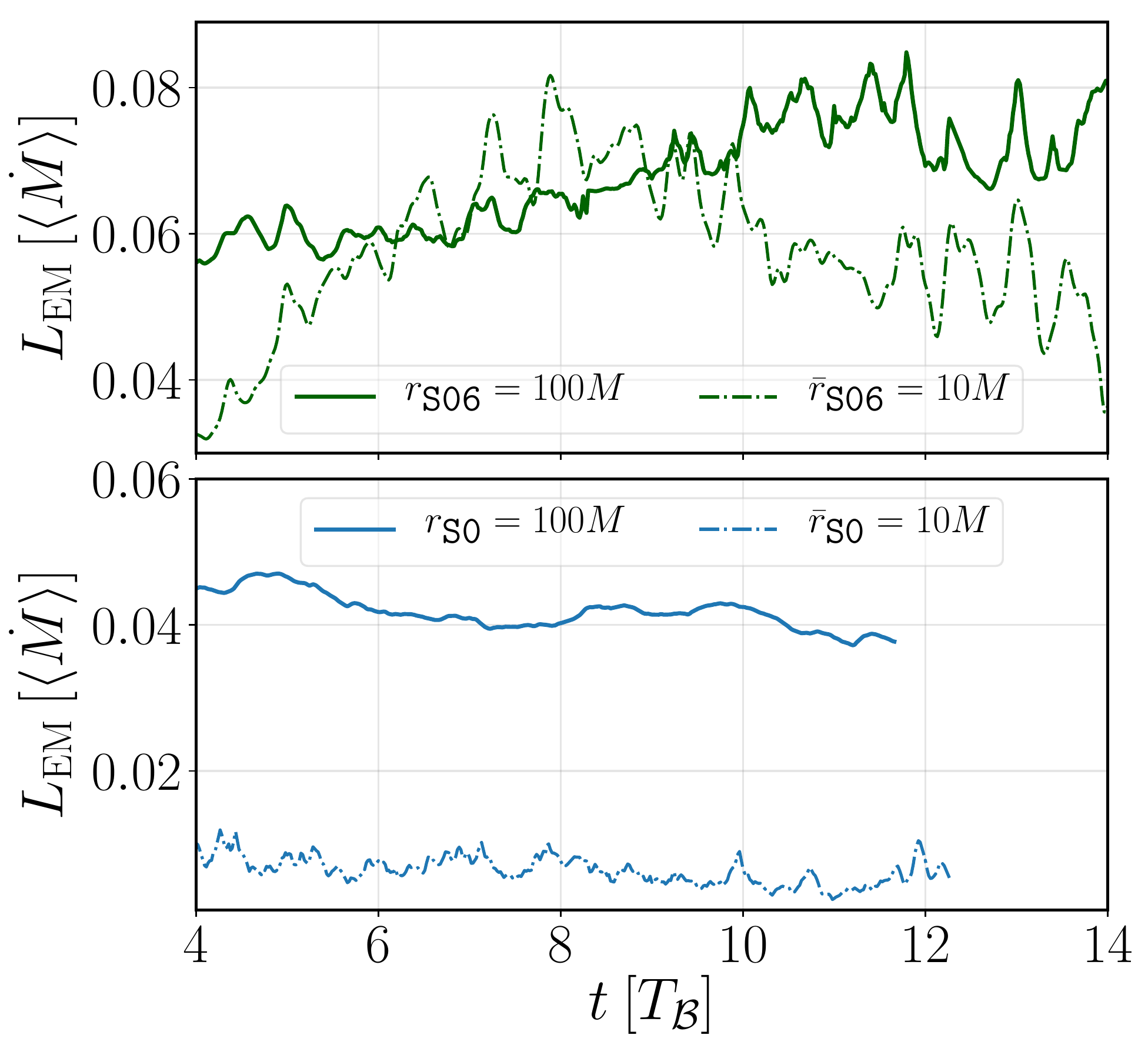}
  \caption{Evolution of the total Poynting flux measured in the BH frame (dashed lines) and in the (inertial) center of mass frame at $100 ~M$(solid lines) for both \ssim{} and \nossim{}. For the center of mass fluxes, we use the retarded time $t-r/\langle v \rangle$ to account for the delay.}
  \label{fig-poyntingcm-t}
\end{figure}

\firstrev{In Figure \ref{fig-poyntingcm-t}, we plot this quantity for both \ssim{} and \nossim{} as a function of retarded time $t-r/\langle v \rangle$, where $\langle v \rangle$ is the mean velocity of the outflow. We also plot the sum of the Poynting fluxes around each mini-disk, measured in the BH comoving frame. We notice that the fluxes in \nossim{} are in average five times larger in the center of mass frame compared with the BH frame, suggesting that there is a kinematic contribution to $L_{\rm EM}$ from the orbital motion of the black holes and possibly from the circumbinary disk. The luminosities in \ssim{}, on the other hand, differ between frames only by tens of percent, suggesting that the rotation of the black hole dominates the extraction of EM energy here \footnote{We have also checked the Poynting luminosity as a function of radius and we observed that, in average, it has varations of around $30 \%$ far from the source, which is expectable as we lose resolution and the outflow interacts with the atmosphere floor of the simulation.}. During our simulations, the speed of the black holes remains fairly constant at $v\sim 0.1$ but, closer to the merger, the orbital speed might enhance significantly the electromagnetic luminosity, as seen e.g. by \citep{palenzuela2010dual, Farris2012, kelly2020electromagnetic}.} 

\begin{figure}[htb]
  \includegraphics[width=\columnwidth]{\figfolder/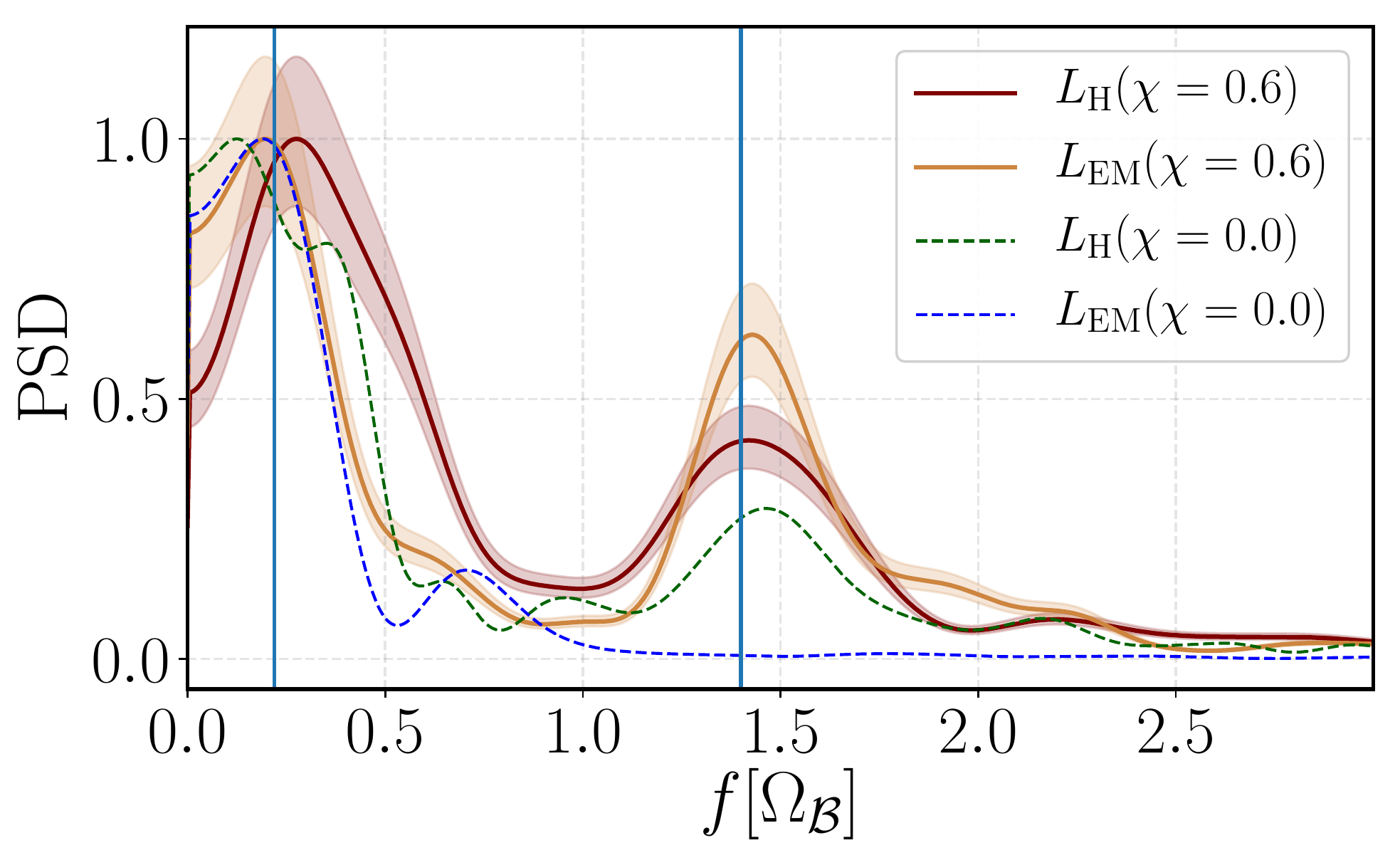}
  \caption{Power spectral density of the hydro ($L_{\rm H}$) and EM luminosities ($L_{\rm EM}$) for \ssim{} (thick lines) and \nossim{} (dashed lines) at $100~ M$ using a Welch algorithm with a Hamming window size and a frequency of $10~M$. The confidence intervals at $3 \sigma$ are shown as shadowed areas for \ssim{}. The two main peaks are given by twice the beat frequency, $2 \Omega_{\rm beat} = 1.4 \Omega_{\rm bin}$, and the lump accretion periodicity $\sim 0.22\Omega_{\rm bin}$}
  \label{fig-psd_lum}
\end{figure}

In both \ssim{} and \nossim{} the EM luminosities are variable, and in both a Fourier power spectrum reveals a periodic modulation at the frequency of the circumbinary disk's inner edge, i.e., the ``lump" frequency.  However, in \ssim{} there is an additional modulation of similar amplitude at twice the beat frequency, proving it is tied closely to accretion, see Figure \ref{fig-psd_lum}.

\begin{figure*}[ht!]
\begin{center}
  \includegraphics[width=\columnwidth]{\figfolder/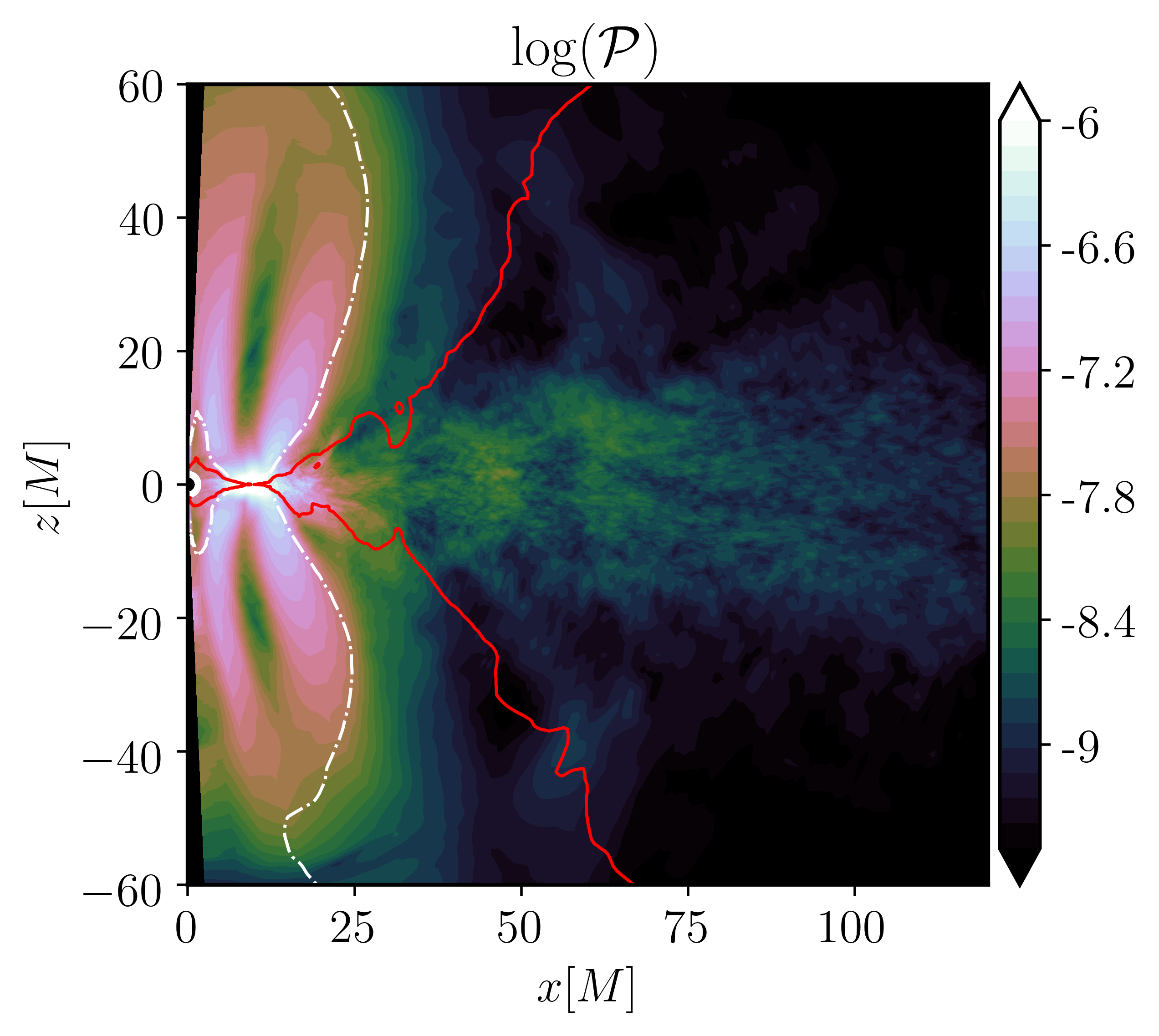}
  \includegraphics[width=\columnwidth]{\figfolder/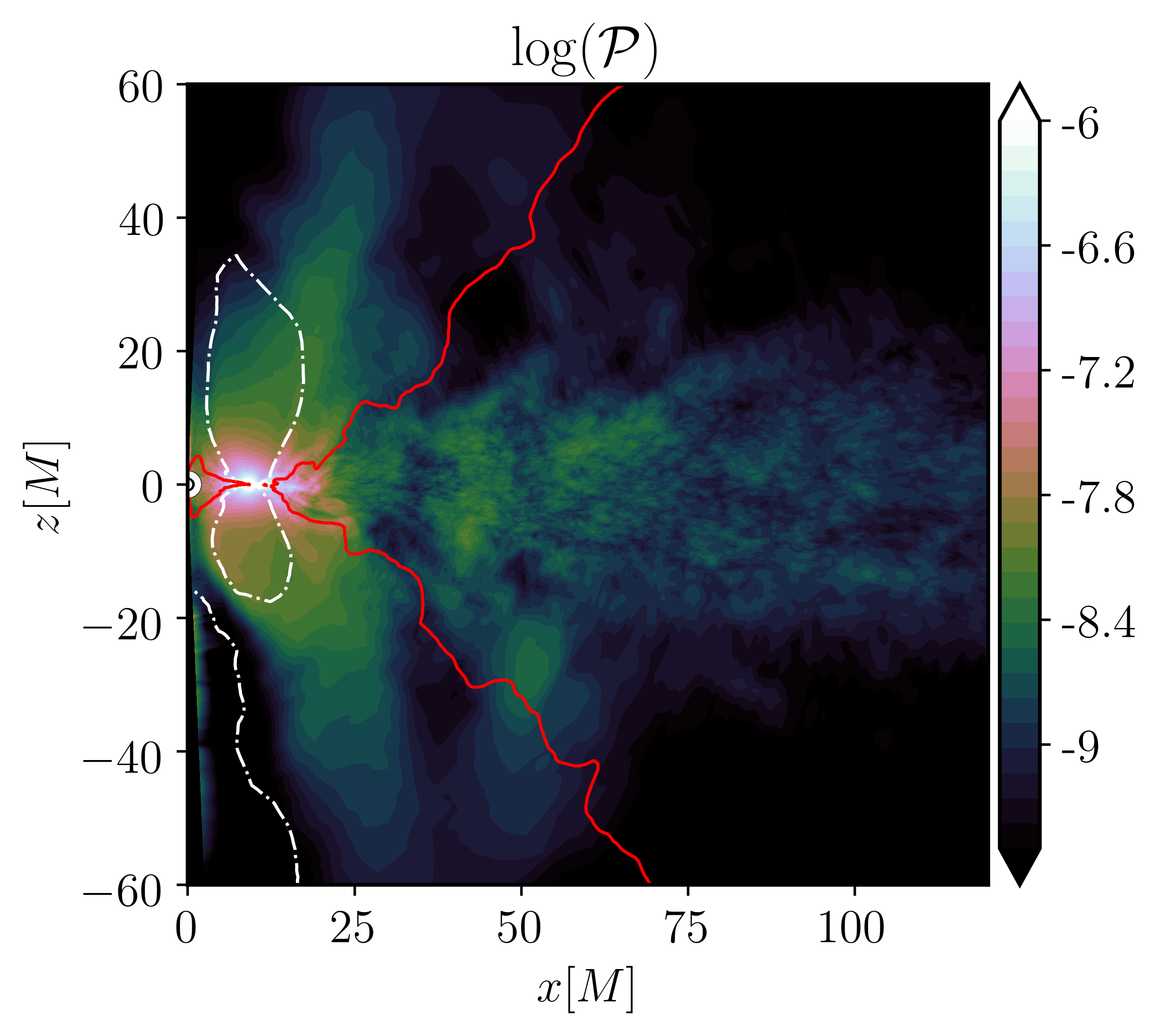}
  \caption{Meridional plot of a time average Poynting scalar for BH$_1$ in \ssim{} (left) and in \nossim{} (right). The black hole is at $x \sim 10M$ and the center of mass is at $x=0M$. The red lines represent the division between bound and unbound material, while the dot-dashed white lines represent the magnetically dominated material.}
  \label{fig-ponynt-pol}
\end{center}
\end{figure*}

To further study the spatial distribution of the electromagnetic flux, in Figure \ref{fig-ponynt-pol}, we plot a meridional slice of the Poynting scalar $\mathcal{P}:= \mathcal{S}^{\bar{i}} \mathcal{S}_{\bar{i}}$ for both \ssim{} and \nossim{},  averaged in time over half an orbit at $4000 ~ M$. The dotted-dashed white lines are defined by $b^2/\rho=1$, containing the regions that better approximate the magnetically dominated region, while the red solid lines are defined by the region where $h u_t =-1$, where the fluid becomes unbound. As expected, \ssim{} has a much more prominent Poynting jet structure than \nossim{}, which has almost zero jet power. The poloidal distribution of $\mathcal{P}$ in \ssim{} around the black hole has a parabolic shape, with most of the flux being emitted at mid-latitudes. Each individual jet shape is similar to those found around single BHs \citep{nakamura2018}. In both cases, we can also notice the strong (bound) Poynting fluxes generated by magnetic stresses in the disk at the equatorial plane.

\begin{figure}[htb]
  \includegraphics[width=\columnwidth]{\figfolder/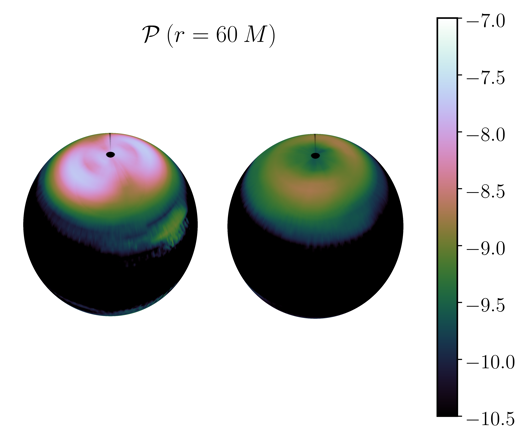}
  \caption{Time average of Poynting scalar $\mathcal{P}$ projected on a sphere of radius 60 M for spinning (left sphere) and non-spinning (right sphere) for unbound elements of fluid.}
  \label{fig-poyntingcm}
\end{figure}

\begin{figure}[htb]
  \includegraphics[width=\columnwidth]{\figfolder/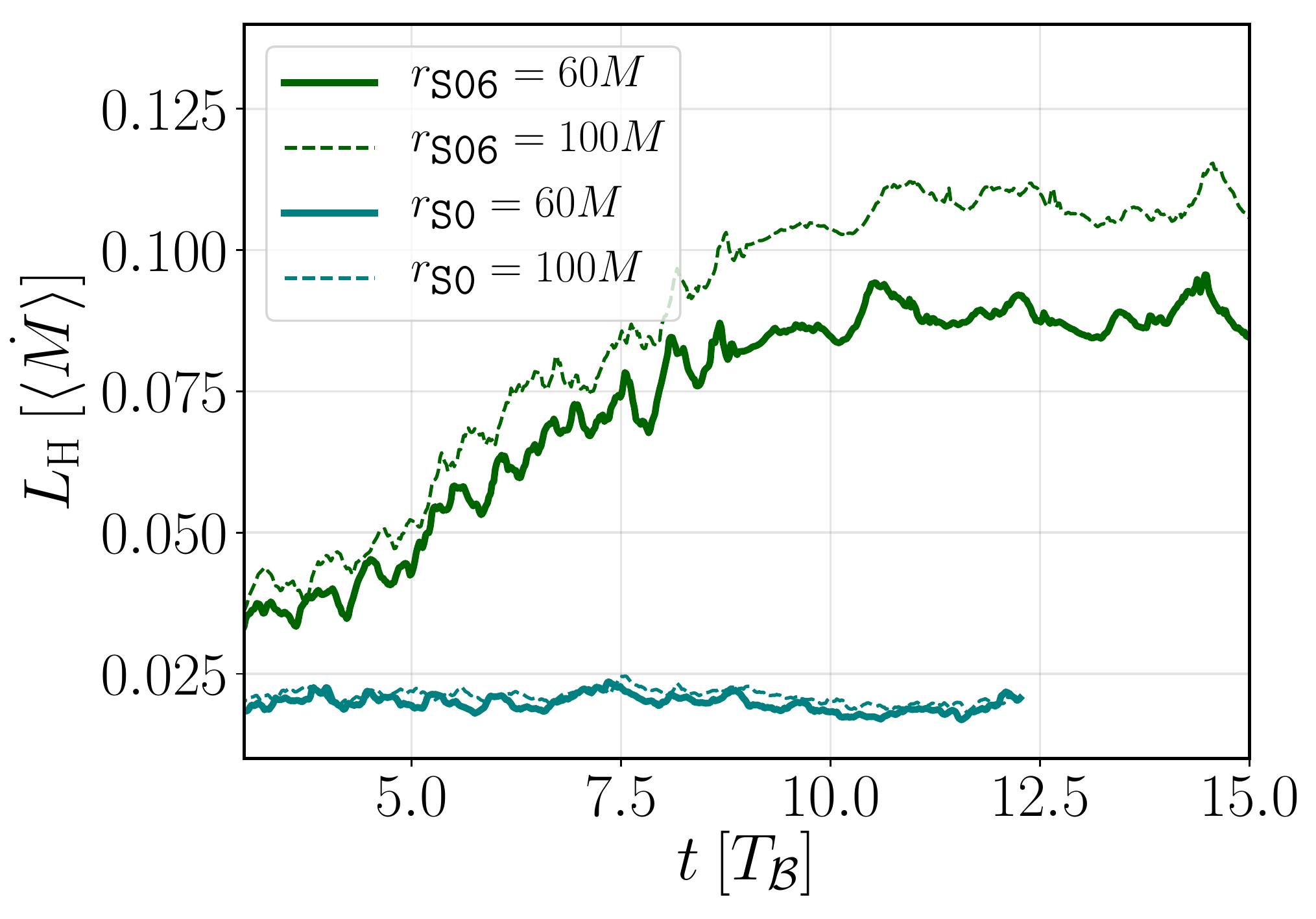}
  \caption{Hydro luminosity as a function of time measured in the center of mass frame for \ssim{} (green lines) and \nossim{} (blue lines) at different radii.}
  \label{fig-koutflows}
\end{figure}

In Figure \ref{fig-poyntingcm}, we plot $\mathcal{P}$, averaged over half an orbit in the corotating frame of the binary, for a sphere of radius $r=60 ~M$ at the center of mass.  In \ssim{} the Poynting flux has a double cone structure that extends to larger polar angles than does the Poynting flux in \nossim{}, likely because of interaction between the two jets.  On the other hand, in \nossim{}, the Poynting flux is distributed on a uniform ring around the axis of the binary.  Unfortunately, our simulation coordinates require a cutout in the grid covering the polar axis running through the center of mass. This prevents an accurate study of the interaction of the jets. Nonetheless, because this cutout is small ($2 ^{\circ}$) compared to the angular size of the jets, we are still able to pick out important features.

Besides the electromagnetic fluxes, there are also hydrodynamical fluxes from the system. We define the hydro luminosity as the integral of the energy flux component of the hydrodynamic stress tensor minus the contribution of the rest-mass energy flux at that radius in order to get the `usable' energy flux \citep{HK06}:
\begin{equation}
L_{\rm H} (t,r) := \Big( \oint_r dA \:(T_{\rm H})^r_t \Big ) - \dot{M}_{\rm jet}
\label{eq-hydroflux}
\end{equation}

In Figure~\ref{fig-koutflows}, we plot the hydro luminosity in the center of mass frame as a function of time for both \nossim{} and \ssim{}, at different radii, normalized by the averaged value of the accretion rate,  so that these luminosities, too, can be described in the language of rest-mass efficiency. In the integral over flux (Eqn.~\eqref{eq-hydroflux}), we include only fluid elements that are both unbound according to the Bernoulli criterion ($-h u_t > 1$) and moving outward ($u^r>0$) (cf. \cite{dVHKH05}). \firstrev{Like the EM luminosity in S06, but not S0, the hydro luminosities are modulated for both spin cases at the ``lump" frequency and at twice the beat frequency. \secondrev{Figure~\ref{fig-psd_lum} shows the Fourier power spectrum for the spinning case. We observe in Figure~\ref{fig-koutflows} a secular growth of the hydro fluxes in \ssim{} while in \nossim{} they remain rather constant, with an average efficiency of $\sim 2 \%$.  Also like the EM case, the hydro energy flux is considerably greater in S06 than S0, but by a factor $\sim 5$.  Such a contrast resembles the differences between the hydro efficiencies measured in spinning and non-spinning single BH simulations \citep{HK06}.}}

\secondrev{We caution, however, that the luminosities measured at $100~M$ may not be the luminosities received at infinity. Energy can be easily converted from EM to hydro or vice versa.  Here we quote the values at $100~M$ because they are the largest we can measure within our grid.  Nonetheless the comparison between S06 and S0 demonstrates clearly that spinning BBH are much more efficient at creating coherent outflows carrying energy both electromagnetically and hydrodynamically than non-spinning BBH.}

\section{Discussion}
\label{sec-discussion}

Although a few candidates have been identified, the existence of supermassive black hole binaries has not been confirmed. The direct detection of their gravitational waves by LISA or pulsar timing arrays (PTA) remains at least a decade into the future. Nevertheless, upcoming wide-field surveys such as the Vera C. Rubin Observatory, SDSS-V, and DESI, may discover many SMBBH candidates through their electromagnetic emission. 

In order to confirm the presence of a SMBBH, we need to build accurate models and predictions of their electromagnetic signatures. Our GRMHD simulations will be useful for this purpose: as a next step, in \cite{gutierrez2021}, we use these simulations to extract light curves and spectra using ray-tracing techniques \citep{Noble07,dAscoli2018} with different radiation models and different masses. The results in this paper constitute the foundations to interpret the underlying physics of those predictions.

Circumbinary and mini-disk accretion onto an equal-mass binary system has been largely studied in the past in the context of 2D $\alpha-$viscous simulations. These simulations are particularly good for analyzing the very long-term behavior of the system, evolving sometimes for $1000$ orbits. Close to the black holes and at close separations, however, the inclusion of 3D MHD and accurate spacetime dynamics becomes necessary in order to describe the proper mechanisms of accretion and outflow. 2D $\alpha-$disk simulations are not able to include spin effects and most of them do not include GR effects (see, however, \cite{RyanMacFadyen17}). On the other hand, in this work we analyze the balance of hydro accretion from the circumbinary streams and conventional accretion from the internal stresses of the mini-disk; to properly model the latter, we need MHD. Moreover, the presence of a proper black hole, and its horizon, makes the accretion processes entirely self-consistent without adding adhoc sink conditions as used in Newtonian simulations (see,  however, \cite{dittmann2021}). Finally, 3D MHD simulations are necessary to model magnetically-dominated regions and jets. The connection of the accretion and the production of electromagnetic luminosity was one of the main motivations of this work, and impossible to analyze in 2D hydro simulations.

Recently, \cite{paschalidis2021minidisk} presented GRMHD simulations of a system similar to the one analyzed in this paper: equal-mass, spinning binary black holes approaching merger. It is then interesting to compare our results and highlight the differences with their model and analysis. In their paper, they use a slightly higher spin value ($\chi = 0.75$) and explore different spin configurations, including antialigned and up-down directions with respect to the orbital angular momentum. Their system has different thermodynamics than ours, using an ideal-gas state equation with $\Gamma=4/3$ and no cooling. Their focus is on the mass budget of the mini-disk (as in \cite{Bowen2019}) and the electromagnetic luminosity when spin is included. They report that spinning black holes have more massive mini-disks and the electromagnetic luminosity is higher, with quantitative measures similar to what we find in this paper. 

\firstrev{In our work, we analyze in great detail, for the first time, the accretion mechanisms onto the mini-disk and their connection to the circumbinary disk. We show that the BHs accretes in two different ways: through direct plunging of the stream from the lump's inner edge (that dominates the accretion), and through `conventional' stresses of the circular component orbiting the mini-disk. This is qualitatively different than single BHs disks and a direct consequenece of the short inflow time determined by $r_{\rm ISCO} / r_{\rm trunc}$; for larger separations and higher spins, we expect mini-disks to behave closer to conventional single BH disks}. Our simulations also differ significantly in the grid setup and initial data. We start our simulations with an evolved circumbinary disk snapshot, taken from \cite{Noble12}, which is already turbulent and presents a lump (starting the simulation from a quasi-stationary torus, the lump appears after $\sim 50$ orbits at these seperations, once the inner edge has settled). This is very important to accurately describe the periodicities of the system given by the beat frequency, which is set by the orbital motion of the lump. \firstrev{These quasi-periodicities might be different if the thermodynamics change, e.g. if there is no cooling, although currently there are no sufficiently long 3D GRMHD simulations of circumbinary disks exploring this.} Interestingly, we found that the Poynting flux is also modulated by the beat frequency. For BBH approaching merger, this constitutes a possible independent observable if this periodicity is translated to jet emission. As expected, for spinning BHs, we also found more powerful Poynting fluxes, in agreement with \cite{paschalidis2021minidisk}.

With our careful analysis of the accretion onto the mini-disks, we show that a disk-like structure survives for longer as the binary shrinks when the black holes have spin. Further explorations with higher spins will show how far these structures survive very close to merger.

\section{Conclusions}
\label{sec-conclusions}

We have performed a GRMHD accretion simulation of an equal-mass binary black hole with aligned spins of $a= 0.6 ~ M_{\rm BH}$ approaching merger. We have compared this simulation with a previous non-spinning simulation of the same system, analyzing the main differences in mini-disk accretion and the variabilities induced by the circumbinary disk accretion. Our main findings can be summarized as follows:

\begin{itemize}

\item Mini-disks in \ssim{}, where BHs have aligned spins $\chi=0.6$, are more massive than in \nossim{}, where BHs have zero spins, by a factor of two. The mass and accretion rate of mini-disks have quasi-periodicities determined by the beat frequency in both simulations (see Section \ref{sec-mdot}).

\item The material in the mini-disk region can be separated into two components of relatively high and low angular momentum. The low angular momentum component mostly plunges directly from the lump edge, forming a strong single-arm stream. The (supra) Keplerian angular momentum component of the fluid is determined by the size of the ISCO and the truncation radius. We have shown that most of the mini-disk mass in \nossim{} is sub-Keplerian while in \ssim{} most of the material follows closely the Keplerian value up to the end of the evolution when it becomes comparable to the low angular momentum component. For binary parameters as in this simulation, we have also predicted a critical value of $\chi \sim 0.45$ for which most of the mass will have low angular momentum relative to the ISCO.

\item The accretion rates at the horizon in \ssim{} and \nossim{} are very similar through the evolution (see Fig. \ref{fig-accrate}), as they are dominated by the plunging material from the lump stream (see Figure \ref{fig-mdot-am}).

\item In \ssim{} a jet-like structure is formed self-consistently around each BH (see Figures \ref{fig-rhp} and \ref{fig-ponynt-pol}). Due to the black hole spin, there is a well-defined Poynting flux (see Figure \ref{fig-ponynt-pol}, left panel) with an efficiency of $\eta \sim 8 \%$ (see Figure \ref{fig-efficiency}).  On the other hand, in \nossim{} the efficiency is closer to $\eta \sim 4 \%$, and its Poynting flux is more homogeneous in space (see Figure \ref{fig-ponynt-pol}, right panel).

\item The time evolution of the Poynting flux is modulated by the quasi-periodicity of the accretion, determined by the beat frequency. In the spinning case, the fluxes in the comoving frame grow and start decreasing at $7.5 T_{\mathcal{B}}$, while in the non-spinning these remain fairly constant.

\end{itemize}

\section*{Acknowledgments}

We thank Eduardo Guti\'errez, Vassilios Mewes, Carlos Lousto, and Alex Dittman, for useful discussions.
L.~C., F.~L.~A, M.~C., acknowledge support from AST-2009330, AST- 754 1028087, AST-1516150, 
PHY-1707946. 
L.~C also acknowledges support from a CONICET (Argentina) fellowship. 

S.~C.~N. was supported by AST-1028087, AST-1515982 and OAC-1515969, 
and by an appointment to the NASA Postdoctoral Program at the Goddard Space Flight Center administrated by USRA through a
contract with NASA. J.H.K. was supported by AST-1028111, PHY-1707826, and AST-2009260
D.~B.~B. is supported by the US Department of Energy through the Los Alamos 
National Laboratory. Los Alamos National Laboratory is operated by Triad National 
Security, LLC, for the National Nuclear Security Administration of U.S. 
Department of Energy (Contract No. 89233218CNA000001).

Computational resources were provided by the Blue Waters sustained-petascale computing NSF Projects No. OAC-1811228 and No. OAC-1516125”, 
and replace with "Computational resources were provided by the TACC’s Frontera supercomputer allocation No. PHY-20010 and AST-20021
Additional resources were provided by the RIT's BlueSky and Green Pairie Clusters   
acquired with NSF grants AST-1028087, PHY-0722703, PHY-1229173 and PHY-1726215.

The views and opinions expressed in this paper are those of the authors 
and not the views of the agencies or US government.

\hfill \break
\bibliography{bhm_references}

\end{document}